\documentclass[preprint,12pt]{elsarticle}

\usepackage{amssymb}
\usepackage{graphicx}
\graphicspath{{PEBGLS_framework_journal_paper_resource/}}

\usepackage{array}
\usepackage{algorithm}
\usepackage{algorithmic}
\usepackage{multirow}
\usepackage{subfigure}
\usepackage{amsmath,amsfonts}

\usepackage{color}

\journal{}%Applied Soft Computing}

\begin{document}

\begin{frontmatter}



\title{A New Cooperative Framework for Parallel Trajectory-Based Metaheuristics}


\author[add_xjtu]{Jialong Shi} \ead{shi.jl@outlook.com}

\author[add_cityu,add_cityu_shenzhen]{Qingfu Zhang} \ead{qingfu.zhang@cityu.edu.hk}

\address[add_xjtu]{School of Mathematics and Statistics, Xi'an Jiaotong University, Xi'an, China}

\address[add_cityu]{Department of Computer Science, City University of Hong Kong, Hong Kong}

\address[add_cityu_shenzhen]{The City University of Hong Kong Shenzhen Research Institute,
Shenzhen, China}

\begin{abstract}
  In this paper, we propose the Parallel Elite Biased framework (PEB framework) for parallel trajectory-based metaheuristics. In the PEB framework, multiple search processes are executed concurrently. During the search, each process sends its best found solutions to its neighboring processes and uses the received solutions to guide its search. Using the PEB framework, we design a parallel variant of Guided Local Search (GLS) called PEBGLS. Extensive experiments have been conducted on the Tianhe-2 supercomputer to study the performance of PEBGLS on the Traveling Salesman Problem (TSP). The experimental results show that PEBGLS is a competitive parallel metaheuristic for the TSP, which confirms that the PEB framework is useful for designing parallel trajectory-based metaheuristics.
\end{abstract}


\end{frontmatter}


\section{Introduction}
Metaheuristics are often used to find nearly optimal solutions of hard optimization problems within a reasonable amount of time. There are two main categories of metaheuristics~\cite{nesmachnow2014overview,jaradat2016effect}: trajectory-based metaheuristics and population-based ones. A trajectory-based metaheuristic iteratively improves a single solution and forms a search trajectory in the solution space. Examples of trajectory-based metaheuristics include Simulated Annealing (SA), Tabu Search (TS), Iterated Local Search (ILS) and Guided Local Search (GLS)~\cite{voudouris1999guided}. In population-based metaheuristics, a population of solutions is processed by several operators at each iteration (generation). The members of the population are replaced by new ones so that the solution space can be explored. Genetic Algorithm (GA), Ant Colony Optimization (ACO), Particle Swarm Optimization (PSO) and Artificial Bee Colony (ABC)~\cite{karaboga2011novel} are some widely-used population-based metaheuristics.

\begin{figure}%[H]
  \centering
  \includegraphics[width=\linewidth]{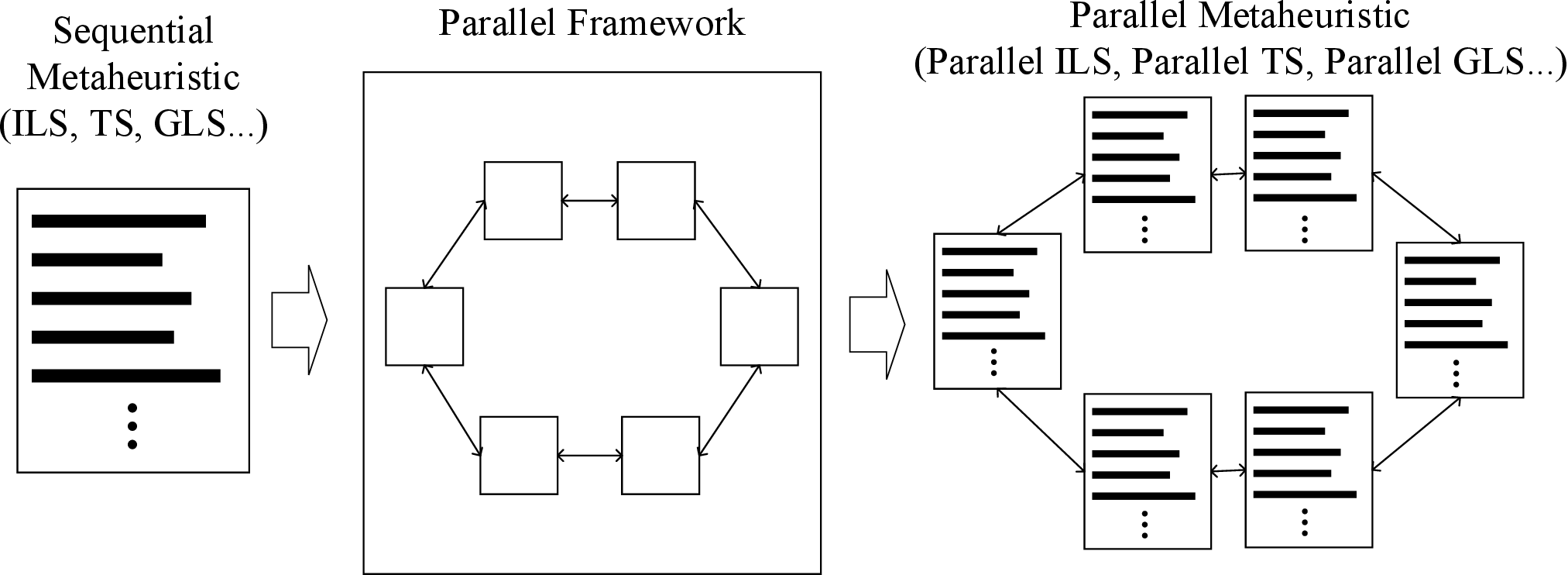}\\
  \caption{Following a parallel framework, one can design the parallel variants of different sequential metaheuristics}\label{fig:famework_alg}
\end{figure}

With the increasing popularity of multi-processor and multi-core platforms, parallelism has become ubiquitous in today's computer technologies. Hence, parallel metaheuristics have attracted a lot of research effort. Here we argue that designing a \emph{parallel framework} is more essential than designing a parallel metaheuristic. A parallel framework is a universal model for designing the parallel variants of a certain kind of metaheuristics. As sketched in Figure~\ref{fig:famework_alg}, a parallel framework defines how multiple metaheuristic processes cooperate with each other and one can apply different sequential metaheuristics to this framework to design different parallel metaheuristics. In this paper we propose a parallel framework which can be used to design the parallel variants of trajectory-based metaheuristics including TS, GLS, etc.

In~\cite{shi2017parallel}, we have proposed a Parallel Elite Biased Tabu Search (PEBTS) algorithm for the Unconstrained Binary Quadratic Programming (UBQP) problem. We generalize the parallelism strategies of PEBTS and propose the Parallel Elite Biased framework (PEB framework) in this paper. Using the PEB framework, we design a new parallel variant of GLS, called Parallel Elite Based GLS (PEBGLS). Extensive experiments are conducted to study the performance and behavior of PEBGLS using the symmetric Traveling Salesman Problem (TSP) as the test suite. We hope that our study can provide a new possible direction for designing parallel metaheuristics.

This paper is structured as follows. Section~\ref{sec:related_works} reviews the related works. In Section~\ref{sec:parallel_framework}, the PEB framework is presented and discussed. In Section~\ref{sec:PEBGLS} we design a parallel variant of GLS following the PEB framework. Section \ref{sec:epm_study} gives the experimental studies on the TSP. Section \ref{sec:conclusion} concludes this paper.

\section{Related Works}\label{sec:related_works}
The existing parallelism strategies of trajectory-based metaheuristics fall into the following two categories.
\begin{itemize}
  \item \emph{Low-level parallelism} or \emph{acceleration strategy}. This strategy aims at speeding up a sequential metaheuristic. It does not change the behavior of the original sequential metaheuristic. The implementations of this strategy usually use the master-slave topology. The master controls the main procedure, dispatching tasks to the slaves. The tasks can be evaluating moves, or evaluating partial solutions. After collecting and integrating the results returned by the slaves, the master proceeds to the subsequent procedure.
  \item \emph{High-level parallelism} or \emph{multiple search strategy}. In this strategy, multiple search processes are executed simultaneously. Each process makes a unique trajectory in the search space. The heuristic methods and initial solutions of the search processes can be the same or different. They may run independently and communicate at the end to identify the best overall solution, or they may exchange useful information during the search. %In the latter case, the behavior of the parallel metaheuristic is different from the corresponding sequential one.
\end{itemize}
A number of high-level parallel trajectory-based metaheuristics have been proposed for various problems. Table~\ref{tbl:liter_review} summarizes the related works on parallel trajectory-based metaheuristics.

\begin{table}%[!h]
\caption{Literature Review}
\centering
\label{tbl:liter_review}
\resizebox{1\linewidth}{!}{
\begin{tabular}{ >{\raggedright}p{120pt} >{\raggedright}p{100pt}  >{\raggedright}p{90pt}  >{\raggedright}p{120pt}  >{\raggedright}p{80pt}  >{\raggedright\arraybackslash}p{120pt} }%>{\raggedleft}p{40pt} | >{\centering\arraybackslash}p{90pt}
\hline
\textbf{Related Work} & \textbf{Algorithm} & \textbf{Problem} & \textbf{Information Type} & \textbf{Information Exchanging Method} & \textbf{Information Utilizing Method}\\
\hline
Garcia-Lopez et al. 2002~\cite{garcia2002parallel} & Parallel VNS & P-median problem & Best found solutions & Centralized & Restart\\
\hline
Bortfeldt et al. 2003~\cite{bortfeldt2003parallel} & Parallel TS & Container loading problem & Best found solutions & Distributed & Restart\\
\hline
Attanasio et al. 2004~\cite{attanasio2004parallel} & Parallel TS & \scriptsize{Dynamic multi-vehicle dial-a-ride problem} & Best found solutions \& Visiting frequencies & Distributed & Restart \& Refer to the frequencies \\
\hline
Banos et al. 2004~\cite{banos2004parallel} & Parallel SA-TS & Graph partitioning & Best found solutions & Distributed & Restart \\
\hline
Crainic et al. 2004~\cite{crainic2004cooperative} & Parallel VNS & P-median problem & Best found solutions & Centralized & Restart \\
\hline
Blazewicz et al. 2004~\cite{blazewicz2004parallel} & Parallel TS & 2-dimensional cutting & Best found solutions & Centralized & Restart \\
\hline
Le Bouthillier and Crainic 2005~\cite{le2005cooperative} & \scriptsize{Parallel TS, EA \& Post-optimization} & VRPTW & Elite solutions & Centralized & Path-relinking \\
\hline
Le Bouthillier, et al. 2005~\cite{le2005guided} & \scriptsize{Parallel TS, EA, Post-optimization \& Pattern identification} & VRPTW & Elite solutions \& Solution attributes & Centralized & Restart \& Fix or prohibit the attributes\\
\hline
Talbi and Bachelet 2006~\cite{talbi2006cosearch} & \scriptsize{COSEARCH (GA, Kick Operator \& TS)} & QAP & Elite solutions \& Global frequencies & Centralized & Restart \& Refer to the frequencies \\
\hline
Fischer and Merz 2005~\cite{fischer2005distributed} 2007~\cite{fischer2007embedding} & \scriptsize{Parallel Chained Lin-Kernighan} & TSP & Best found solutions & Distributed & Restart\\
\hline
Lukasik et al. 2007~\cite{lukasik2007parallel} & Parallel SA & Graph coloring problem & Best found solutions & Centralized & Restart \\
\hline
Ribeiro and Rosseti 2007~\cite{ribeiro2007efficient} & Parallel GRASP & 2-path network design problem & Elite solutions & Centralized & Path-relinking \\
\hline
Araujo et al. 2007~\cite{araujo2007exploring} & Parallel GRASP-ILS & \scriptsize{Mirrored traveling tournament problem} & Elite solutions & Centralized & Restart \\
\hline
Aydin and Sevkli 2008~\cite{aydin2008sequential} & Parallel VNS & Job shop scheduling & Best found solutions & Distributed & Restart \\
\hline
Polacek et al. 2008~\cite{polacek2008cooperative} & Parallel VNS & MDVRPTW & Best found solutions & Centralized & Restart \\
\hline
Dos Santos et al. 2009~\cite{dos2009parallel} & \scriptsize{Parallel GRASP, GA \& Q-learning} & TSP & Best found solutions & Centralized & Restart \& Update Q-values table \\
\hline
Luque et al. 2010~\cite{luque2010new} 2011~\cite{luque2011exploring} & Parallel SA & \scriptsize{DNA fragment assembly, MAXSAT \& RND} & Best found solutions & Distributed & Combination operation\\
\hline
Subramanian et al. 2010~\cite{subramanian2010parallel} & ILS-RVND & VRPSPD & Best parameter values & Centralized & Set the parameters \\
\hline
Hung and Chen 2011~\cite{hung2011heterogeneous} & \scriptsize{Parallel Branch-and-Bound method \& TS} & TSP & Best found solutions & Centralized & Restart \\
\hline
Cordeau and Maischberger 2012~\cite{cordeau2012parallel} & Parallel ILS-TS & VRP &  Best found solutions & Distributed & Restart with a probability\\
\hline
Lee et al. 2012~\cite{lee2012parallel} & Harmony search & Task scheduling problem & Elite solutions & Centralized & Restart \\
\hline
Jin et al. 2014~\cite{jin2014cooperative} & Parallel TS & Capacitated VRP & Best found solutions & Centralized & Restart\\
\hline
Hemmelmayr 2015~\cite{hemmelmayr2015sequential} & \scriptsize{Parallel Large Neighborhood Search} & \scriptsize{Periodic Location Routing Problem} & Best found solutions & Centralized & Restart\\
\hline
Iturriaga et al. 2015~\cite{iturriaga2015parallel} & Parallel stochastic LS & \scriptsize{Heterogeneous Computing Scheduling} & Neighboring solutions & Centralized & Evaluate \& report\\
\hline
Lahrichi et al. 2015~\cite{lahrichi2015integrative} & Integrative Cooperative Search & MDPVRP & Solutions \& partial solutions & Centralized & Restart or integrate\\
\hline
Luque and Alba 2015~\cite{luque2015parallel} & Parallel SA & \scriptsize{DNA fragment assembly \& QAP} & Best found solutions & Distributed & Path-relinking\\
\hline
Tosun 2015~\cite{tosun2015performance} & Parallel GA \& TS & QAP & Elite solutions & Centralized & Crossover \& restart\\
\hline
Wang  et al. 2015~\cite{wang2015parallel} & Parallel SA & VRPSPDTW & Best found solutions & Centralized & Restart \\
\hline
Sousa Filho et al. 2016~\cite{sousa2016parallel} & GRASP-VNS & Bicluster editing problem & Best found solutions & Centralized & Restart \\
\hline
Guzman et al 2016~\cite{guzman2016novel} & Parallel TS \& SA & TSP & Best found solutions & Centralized & Restart\\
\hline
Quan \& Wu 2017~\cite{quan2017design} & Parallel ILS & \scriptsize{Disjunctively Constrained Knapsack Problem} & Elite solutions & Centralized & Restart\\
\hline
Tu et al. 2017~\cite{tu2017spatial} & Parallel ILS & VRP & Subproblems \& solutions & Centralized & run ILS on subproblem\\
\hline
\end{tabular}
}
\end{table}

As shown in Table \ref{tbl:liter_review}, in most of the existing parallel trajectory-based metaheuristics, the parallel processes exchange solutions with each other. The solutions can be the best solutions found so far or some elite solutions. In Table \ref{tbl:liter_review}, the centralized communication method is used by most of the metaheuristics, while some metaheuristics apply the distributed communication method, in which each process only shares information with a limited number of processes. Compared to the centralized communication method, the distributed communication method is more flexible and can be used in massive parallel processing platforms.

In Table \ref{tbl:liter_review}, the information utilizing methods in many existing parallel trajectory-based metaheuristics are denoted as ``restart''. In this method, when a process receives a new solution, it abandons the current solution and restarts from the received one. As a consequence, the information in the current solution is lost. Some works try to overcome this drawback by using path-relinking methods or other combination operators. In their methods, a new solution is generated based on the received solution and the current solution.

\section{Parallel Elite Biased Framework} \label{sec:parallel_framework}
In this section, we propose the Parallel Elite Biased framework (PEB framework) for the designing of parallel trajectory-based metaheuristics with multiple search processes. To design a parallel trajectory-based metaheuristic with multiple search processes, three issues must be addressed:
\begin{itemize}
  \item What information should be exchanged among different processes? (information type)
  \item How should information be exchanged? (communication method)
  \item How should the received information be utilized? (information utilizing method)
\end{itemize}
For the first issue, in the PEB framework, different processes exchange their best found solutions with each other. For the second issue, the PEB framework follows a distributed topology in which each process only sends solutions to their predefined neighbors. For the third issue, the PEB framework applies a novel information utilizing method. In the related works, when a process receives a new solution, it either restarts from the received solution or executes path-relinking to generate a new solution. In the PEB framework, the process continues searching from its current solution, but its search will have bias toward the received solutions. %In such a way, the historical information of each process is used efficiently.

\subsection{Communication Method}
\begin{figure}
  %\centerline{
  \centering
  \subfigure[Bidirectional ring]{
    \label{fig:stru_bir} %% label for first subfigure
    \includegraphics[width=0.35\linewidth]{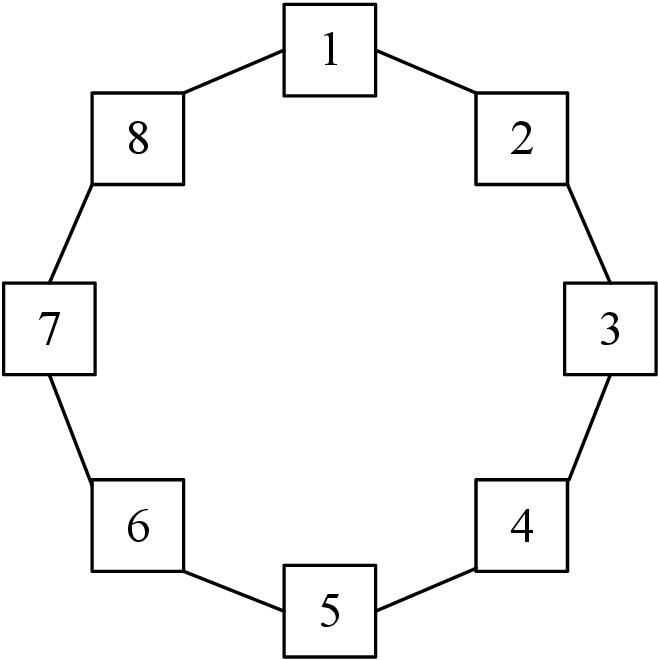}}
  \hspace{0.05\linewidth}
  \subfigure[Torus]{
    \label{fig:stru_tor} %% label for first subfigure
    \includegraphics[width=0.35\linewidth]{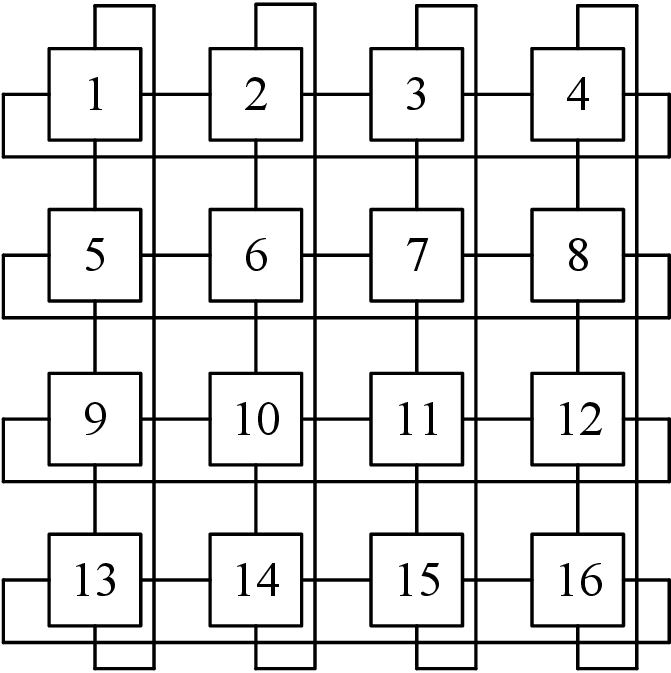}}
  %}
  \caption{Different parallel topologies}\label{fig:2_parallel_stru} %% label for entire figure
\end{figure}

The proposed PEB framework is based on a distributed topology. To reduce the communication load, each process only communicates with a number of neighboring processes. In this paper, we consider two distributed topologies: the bidirectional ring topology and the torus topology. They are two of the most natural distributed topologies and widely used in the area of parallel metaheuristics~\cite{alba2005parallel,crainic2010parallel}. Both topologies support flexible process number. Figure \ref{fig:2_parallel_stru} illustrates the examples of these parallel topologies. In the bidirectional ring topology, each process has two neighbors. For example, in Figure \ref{fig:2_parallel_stru}(a), the neighbors of process 1 are process 2 and process 8. In the torus topology, each process has four neighbors. For example, in Figure \ref{fig:2_parallel_stru}(b) the neighbors of process 1 are process 2, process 5, process 4 and process 13.

A well-designed parallel metaheuristic must control the communication load of the processes. A rigidly synchronous communication strategy may cause heavy communication load and reduce the efficiency of the parallel metaheuristic. The PEB framework follows an asynchronous communication pattern. For each search process in the PEB framework, we denote $s_{hb}$ as the historical best solution found by itself and $S_r$ as the set of solutions received from its neighboring processes. After a given period of time, each process checks whether $s_{hb}$ has changed since the previous sending. If so, it sends the new $s_{hb}$ to its neighbors. Meanwhile each process keeps receiving new solutions from its neighboring processes. Note that, although a process may receive better solutions from its neighbors, it always sends the best solution found by itself to its neighbors. This maintains the diversity of the parallel metaheuristic. This strategy is helpful to reduce the communication load among processes.

\subsection{Information Utilizing Method}
In the PEB framework, each process maintains an elite solution $s_e$, which is the best solution in the set $S_r \cup \{s_{hb}\}$. The PEB framework applies a novel method to utilize the information in $s_e$. Instead of restarting from the $s_e$, each process continues searching from the current solution and its search procedure has bias toward $s_e$. In other words, each process is ``attracted'' by $s_e$. In such way, each process can utilize the information in $s_e$ without abandoning the information in its current solution. In practise, the way to realize the attraction of $s_e$ is decided by users.

\begin{figure}
  %\centerline{
  \centering
  \subfigure[Independent]{
    \label{fig:coop_ind} %% label for first subfigure
    \includegraphics[width=0.4\linewidth]{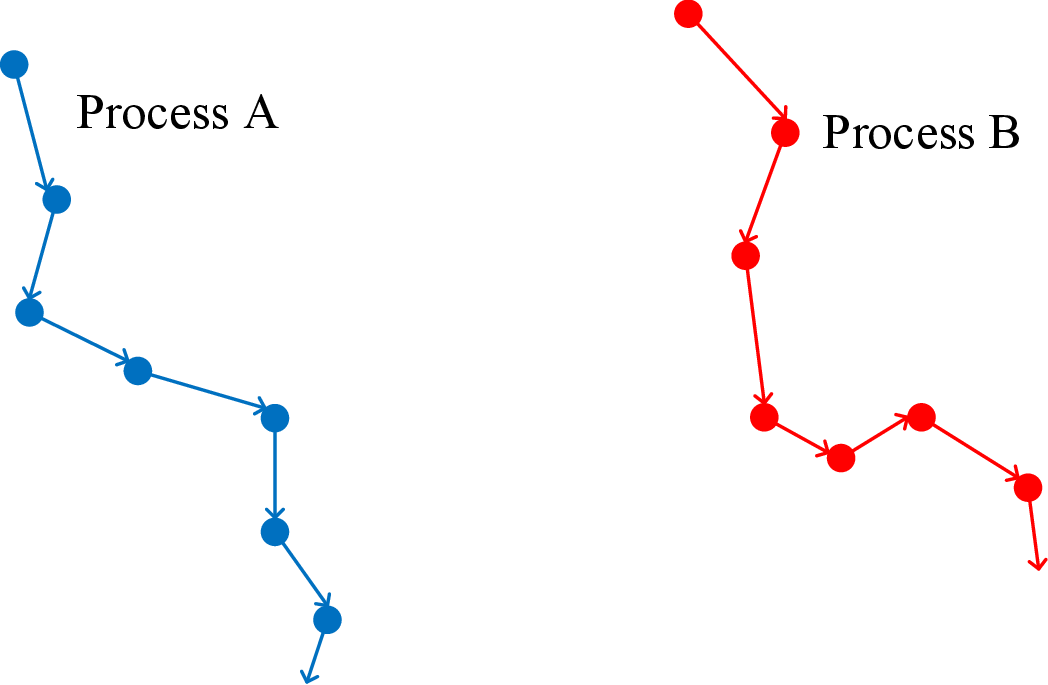}}
  \hspace{0.1\linewidth}
  \subfigure[Restart]{
    \label{fig:coop_res} %% label for first subfigure
    \includegraphics[width=0.4\linewidth]{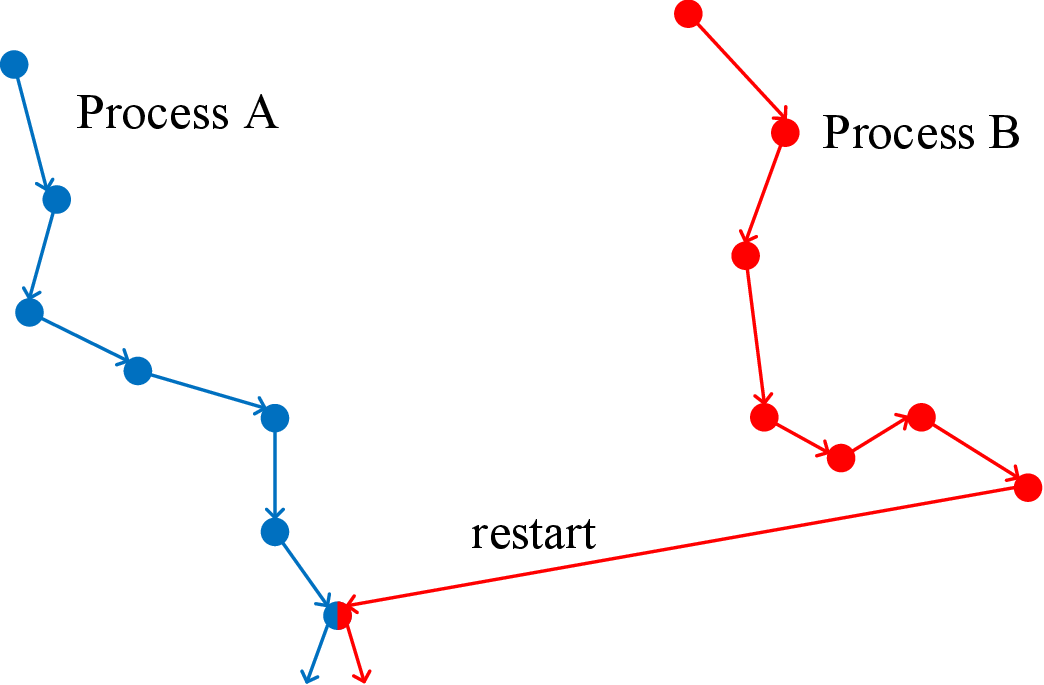}}\\
  \subfigure[Path-Relinking]{
    \label{fig:coop_PR} %% label for first subfigure
    \includegraphics[width=0.4\linewidth]{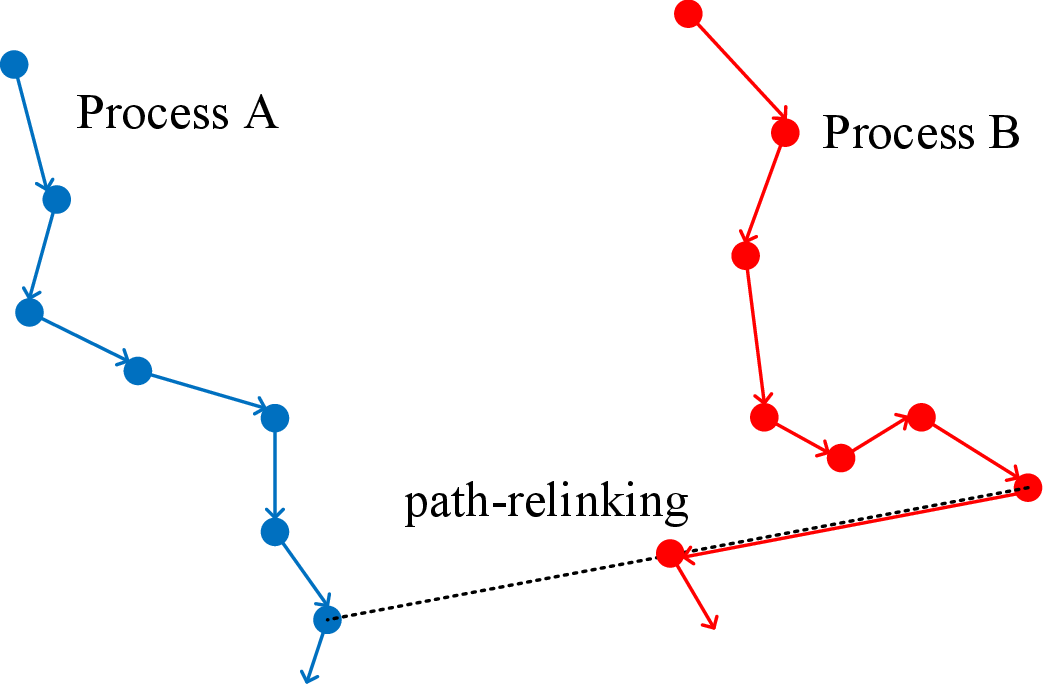}}
  \hspace{0.1\linewidth}
  \subfigure[Elite Biased]{
    \label{fig:coop_EB} %% label for first subfigure
    \includegraphics[width=0.4\linewidth]{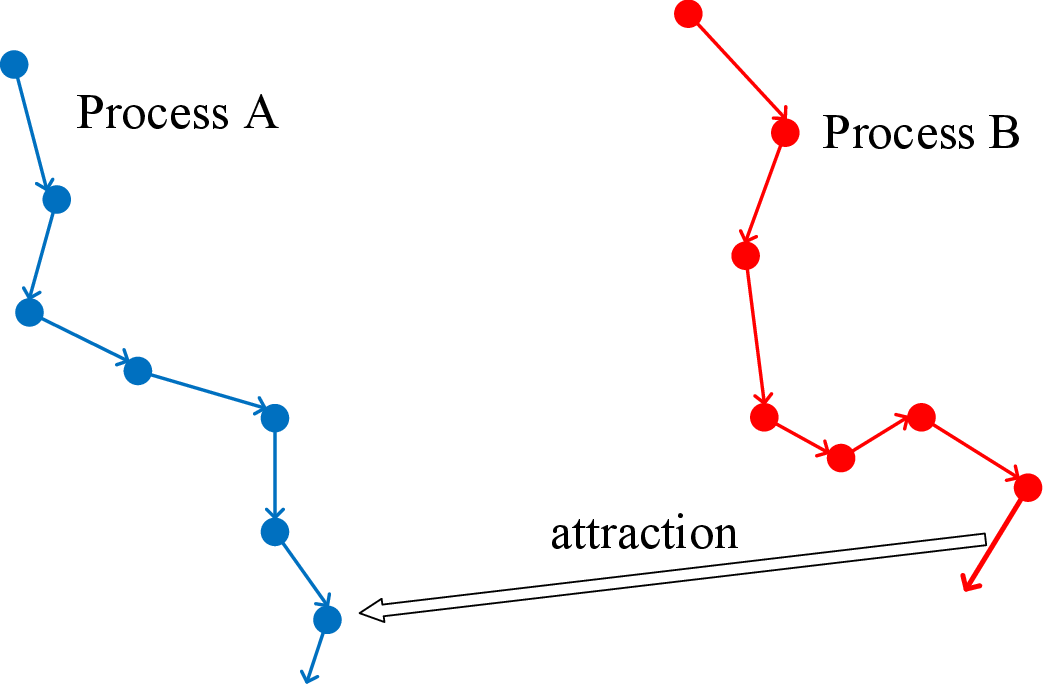}}
  %}
  \caption{Different information utilizing methods}\label{fig:3_coop_method} %% label for entire figure
\end{figure}

To illustrate the information utilizing method of the PEB framework, we give an example in Figure~\ref{fig:3_coop_method}. In Figure \ref{fig:coop_ind}, process A and process B start from different solutions and perform different trajectories. There is no communication between A and B in Figure~\ref{fig:coop_ind}. We compare the information utilizing method of the PEB framework with two widely-used methods: the restart method and the path-relinking method. In the restart method, as shown in Figure~\ref{fig:coop_res}, process B receives a solution from process A and restarts from the received solution. The original solution of process B is abandoned. In the path-relinking method (Figure \ref{fig:coop_PR}), process B generates a new solution using the path-relinking operator based on the received solution and its current solution. Then it proceeds with the search from the resulting solution. The original solution of process B is abandoned too. In the PEB framework (Figure \ref{fig:coop_EB}), process B continues searching from its original solution, but the search direction of process B is attracted by the solution received from process A. The original solution of process B is not abandoned in the PEB framework. This can maintain the diversity of the processes.

\subsection{Pseudocode}
The procedure of each process in the PEB framework is shown in Algorithm \ref{alg:parallel_framework}. In Algorithm \ref{alg:parallel_framework}, the \emph{TryToReceive} procedure always prepares to update the set $S_r$ if it receives new solutions from some neighbors. At the pre-defined time points (e.g. a given number of iterations), the \emph{SelectBestSolution} procedure selects the best one of the set $S_r \cup \{s_{hb}\}$ as $s_e$, and the \emph{SendToNeighbors} procedure sends $s_{hb}$ to all neighbors if $s_{hb}$ has changed since the previous sending. The \emph{EliteBiasedSearch} procedure denotes the combination of the elite biased concept and the original search procedure of a given trajectory-based metaheuristic. The details of the EliteBiasedSearch procedure are decided by users and depend on the mechanism of the metaheuristic users want to parallelize. For example, in PEBTS~\cite{shi2017parallel}, the EliteBiasedSearch procedure is a TS procedure which is influenced by the recorded elite solution $s_e$. In this paper, we propose PEBGLS and its EliteBiasedSearch procedure is a GLS procedure which is presented in Section~\ref{sec:PEBGLS}.

\begin{algorithm}[H]
    \begin{algorithmic}[1]
        \STATE \textbf{initialize}: $s,s_{hb}$
        \WHILE {!StoppingCriterion}
            \STATE $S_r\gets$ TryToReceive()
            \IF{at the pre-defined time points}
                \STATE $s_e\gets$ SelectBestSolution($S_r \cup \{s_{hb}\}$)
                \IF{$s_{hb}$ has been updated since the previous sending}
                    \STATE SendToNeighbors($s_{hb}$)
                \ENDIF
            \ENDIF
            \STATE $\{s,s_{hb}\}\gets$ EliteBiasedSearch($s,s_{hb},s_e$)
        \ENDWHILE
        \RETURN{$s_{hb}$}
    \end{algorithmic}
\caption{Parallel Elite Biased Framework}
\label{alg:parallel_framework}
\end{algorithm}

\section{Designing Parallel Guided Local Search}\label{sec:PEBGLS}
To show the utility of the proposed PEB framework, in this section, we design a parallel variant of Guided Local Search (GLS) following the PEB framework.

\subsection{Guided Local Search}\label{sec:GLS}
GLS is an efficient trajectory-based metaheuristics for combinatorial optimization problems. It iteratively helps a LS procedure to escape from local optima by dynamically adjusting its guide function. We assume that there is a combinatorial optimization problem with solution space $S$ and objective function $g:S\to\mathbb{R}$ to minimize. To apply GLS on this problem, one first needs to define features for candidate solutions in $S$. Each feature has a fixed cost and a penalty. The cost is related to the objective function $g(\cdot)$. The penalty is set to 0 at the beginning and changes during the search. GLS does not use $g(\cdot)$, but the \emph{augmented objective function} $h(\cdot)$ as the guide function of LS:
\begin{equation}\label{eq:h_function}
h(s)=g(s)+ \lambda\sum_{i\in M} p_iI_i(s),
\end{equation}
where $s$ is a candidate solution, $\lambda$ is a pre-defined parameter that controls the penalizing strength, $M$ is the set of all features in the problem, $p_i$ is the current penalty value of feature $i$ and function $I_i(s)$ is an indicator function of whether solution $s$ has feature $i$:
\begin{equation}
I_i(s)= \left\{
    \begin{array}{rl}
        1 &\mbox{ if feature $i$ is in $s$}, \\
        0 &\mbox{ otherwise.}
    \end{array} \right.
\end{equation}

In each iteration, GLS executes a LS using $h(\cdot)$ as the guide function. Once the LS stops at a local optimum $s_*$, GLS adjusts $h(\cdot)$ by increasing the penalties of one or more selected features in $s_*$. To do so, GLS defines the penalizing utility of each feature $i$, $util_i$, as
\begin{equation}\label{eq:def_of_util}
util_i(s_*)=I_i(s_*)\cdot \frac{c_i}{1+p_i},
\end{equation}
where $c_i$ is the cost of feature $i$. GLS selects the features with the highest utility value and increases their penalties by 1. Then a new iteration starts from $s_*$. In (\ref{eq:def_of_util}), the numerator is the cost of feature, which means that features with higher costs are more likely to be penalized and thus low cost features are exploited. The denominator is the accumulated penalty of feature plus 1, which means that the features that has been rarely penalized before have a good chance to be penalized. In such a way, the search explores new regions of the search space.

The pseudocode of GLS is shown in Algorithm \ref{alg:GLS}. The inputs are the objective function $g$, the GLS parameter $\lambda$, the feature set $M$ and the cost of each feature $\{c_i|i \in M\}$.

\begin{algorithm}%[!h]
    \begin{algorithmic}[1]
        \STATE \textbf{input:} $g,\lambda,M, c$
            \STATE $j \gets 0$
            \STATE $s_0 \gets $ random or heuristically generated solution.
            \STATE $s_{hb} \gets s_0$
            \FOR {$i=1\to |M|$}
                \STATE $p_i \gets 0$
            \ENDFOR
            \WHILE {!StoppingCriterion}
                \STATE $h\gets g+\lambda \sum p_iI_i$
                \STATE $\{s_{j+1},s_{hb}\} \gets$ LocalSearch($s_j,s_{hb},h$) %/* including the best-solution-so-far tracking mechanism */
                \FOR {$i = 1\to |M|$}
                    \STATE $util_i \gets I_i(s_{j+1})\cdot c_i/(1+p_i)$
                \ENDFOR
                \FOR {each $i$ such that $util_i$ is maximum}
                    \STATE $p_i\gets p_i + 1$
                \ENDFOR
                \STATE $j\gets j+1$
            \ENDWHILE
        \RETURN{$s_{hb}$}
    \end{algorithmic}
\caption{Guided Local Search}
\label{alg:GLS}
\end{algorithm}

In Algorithm \ref{alg:GLS}, the LS procedure is based on $h(\cdot)$, so GLS needs to track the historical best solution $s_{hb}$ with regard to the original objective function $g$. After each move of LS, GLS checks whether the $g$ value of the new solution is better than that of the recorded best solution, if so, the historical best solution $s_{hb}$ will be updated.

\subsection{Parallel Elite Biased Guided Local Search}
Tairan and Zhang~\cite{tairan2011p} proposed a parallel GLS algorithm called P-GLS-II. However, they first transformed GLS into a population-based metaheuristic, then ran it in a parallel way. To our best knowledge, there is no parallel trajectory-based variant of GLS. Following the proposed PEB framework, we designed a parallel variant of GLS, which is called Parallel Elite Biased GLS (PEBGLS).

\subsubsection{The Attraction of $s_e$}
In the PEB framework, for each process, an elite solution $s_e$ is selected from the set formed by the received solutions and the current historical best solution. Then the search process is attracted by $s_e$. GLS executes LS based on the function $h(\cdot)$ which is augmented by the penalties. Hence the descending nature of the LS will guide GLS to the solutions with less penalties. The proposed parallel variant of GLS aims to reduce the number of penalties imposed on the features that belong to $s_e$ and increase more penalties on the features not belonging to $s_e$. As a result, the search process will be orientated to the search regions near to $s_e$. To achieve this aim, we modified the formula of $util$, i.e. Equation (\ref{eq:def_of_util}). The new formula is:
\begin{equation}\label{eq:modified_util}
util_i(s_*)= \left \{
    \begin{array}{ll}
        I_i(s_*)\cdot {c_i}/{(1+p_i)},&\mbox{ if feature $i$ is in $s_e$};\\
        I_i(s_*)\cdot w\cdot{c_i}/{(1+p_i)},& \mbox{ otherwise,}
    \end{array}
    \right.
\end{equation}
where $w>1$ is a predefined parameter. In Equation (\ref{eq:modified_util}), if a feature is not in $s_e$, its penalizing utility will be multiplied by an extra coefficient $w$. Since $w>1$, features not in $s_e$ will have relatively large $util$ values, so they are more likely to be penalized compared to the features in $s_e$. Hence the penalties imposed on $s_e$ will become relatively small. Due to the descending nature of LS, the search direction of GLS will be attracted by $s_e$.

\subsubsection{The Procedure of PEBGLS}
The procedure of each PEBGLS process is shown in Algorithm \ref{alg:PEBGLS}. In PEBGLS, there is a predefined parameter $U\in\mathbb{N}^+$. Every $U$ iterations, the PEBGLS process updates $s_e$ to the best one of the set $S_r \cup \{s_{hb}\}$ and sends its $s_{hb}$ to all neighbors if $s_{hb}$ has changed since the previous sending. Here $U$ can be used to control the communication load. The inputs of PEBGLS are: objective function $g$, the feature set $M$, the cost of each feature $\{c_i|i \in M\}$ and the user-defined parameters $\{\lambda, w, U\}$.

\begin{algorithm}%[!h]
    \begin{algorithmic}[1]
        \STATE \textbf{input:}$g,M, c,\lambda, w, U$
        \STATE $j \gets 0$
        \STATE $s_0 \gets $ random or heuristically generated solution
        \STATE $s_{hb} \gets s_0$
        \STATE $S_r \gets \emptyset$
        \FOR {$i=1\to |M|$}
            \STATE $p_i \gets 0$
        \ENDFOR
        \WHILE {!StoppingCriterion}
            \STATE $S_r \gets$ TryToReceive()
            \IF{$j\%U==0$}
                \STATE $s_e \gets$ SelectBestSolution($S_r \cup \{s_{hb}\}$)
                \IF{$s_{hb}$ has updated since the previous sending}
                    \STATE SendToNeighbors($s_{hb}$)
                \ENDIF
            \ENDIF
            \STATE $h\gets g+\lambda \sum p_iI_i$
            \STATE $\{s_{j+1},s_{hb}\} \gets$ LocalSearch($s_j,s_{hb},h$) %/* including best-solution-so-far tracking mechanism */
            \FOR {$i = 1\to |M|$}
                \IF {feature $i$ is in $s_e$}
                    \STATE $util_i \gets I_i(s_{k+1})\cdot c_i/(1+p_i)$
                \ELSE
                    \STATE $util_i \gets I_i(s_{k+1})\cdot w \cdot c_i/(1+p_i)$
                \ENDIF
            \ENDFOR
            \FOR {each $i$ such that $util_i$ is maximum}
               \STATE $p_i\gets p_i + 1$
            \ENDFOR
            \STATE $j\gets j+1$
        \ENDWHILE
        \RETURN{$s_{hb}$}
    \end{algorithmic}
\caption{Parallel Elite Biased Guided Local Search}
\label{alg:PEBGLS}
\end{algorithm}

This section shows how we apply the PEB framework to GLS, including how to realize the attraction of $s_e$ in GLS. An example of applying the PEB framework to Tabu Search can be found in \cite{shi2017parallel}. When applying the PEB framework to other kinds of trajectory-based metaheuristics, users need to design unique attraction strategies. Our suggestion is to give priorities to the candidate solutions that are more similar/closer to $s_e$ in each move step of a trajectory-based metaheuristic.

\section{Experimental Studies}\label{sec:epm_study}
In the experimental studies, we tested the performance of PEBGLS on the Traveling Salesman Problem. The TSP is one of the most widely-used benchmarks in the area of combinatorial optimization and it is also one of the most well-known applications of GLS~\cite{voudouris1999guided}.

\subsection{Applying Guided Local Search to the Traveling Salesman Problem}
In the TSP, $G=(V, E)$ is a fully connected graph where $V$ is its node set and $E$ the edge set, $c_e>0$ is the cost of edge $e \in E$. A solution tour $s$ in $G$ is a cycle passing through every node in $V$ exactly once and its cost is defined as:
\begin{equation}
g(s)=\sum_{e \in s}c_e.
\end{equation}
Here $g(\cdot)$ is the objective function of the TSP and the goal of the TSP is to find a tour with the smallest $g$ value. This paper considers the symmetric TSP, where the cost from node $A$ to node $B$ is the same as that from $B$ to $A$. We denote the set of all the feasible tours in $G$ as $S$, which is the solution space of the TSP.

To apply GLS to the TSP, we define that the features are edges in $G$ (i.e. feature set $M$ = edge set $E$) and the features' costs are the costs of the corresponding edges. If a solution tour $s$ contains the edge $e_i$ (i.e. the feature $i$), then $I_i(s)=1$, otherwise $I_i(s)=0$. In this paper, we apply 2-Opt move in GLS, because according to \cite{voudouris1999guided} GLS performs better with 2-Opt, especially when it is combined with the Fast Local Search (FLS) strategy~\cite{bentley1992fast}.

\subsection{Speedup} \label{sec:epm_speedup}
The main purpose of applying parallel metaheuristics is to accelerate the sequential metaheuristics; hence \emph{speedup} is an important metric to measure the performance of parallel metaheuristics. In this section, we measured the speedup of the proposed PEBGLS with different process number on different TSP instances, to study the accelerating ability and scalability of PEBGLS.

The speedup metric compares the runtime of parallel algorithms against the runtime of sequential algorithms. We denote $T_1$ as the runtime of a PEBGLS with only one process and $T_K$ as the runtime of a $K$-process PEBGLS. Note that the runtime measured in parallel algorithms is wall-clock time. Then the speedup $\mathcal{S}_K$ is calculated by:
\begin{equation}\label{eq:speedup1}
\mathcal{S}_K = \frac{E[T_1]}{E[T_K]}.
\end{equation}
where $E[\cdot]$ is the expectation function. If $\mathcal{S}_K<K$, we call it a \emph{sublinear} speedup; if $\mathcal{S}_K=K$, we call it a \emph{linear} speedup; if $\mathcal{S}_K>K$, we call it a \emph{superlinear} speedup. The other widely used metric is \emph{efficiency} $e_K$, which equals $\mathcal{S}_K/K$. Obviously $e_K\geq1$ is desirable.

\begin{figure}
  %\centerline{
  \centering
  \subfigure[$\mathcal{S}_K$ on att532]{
    %\label{fig:speedup1}
    \includegraphics[width=0.32\linewidth]{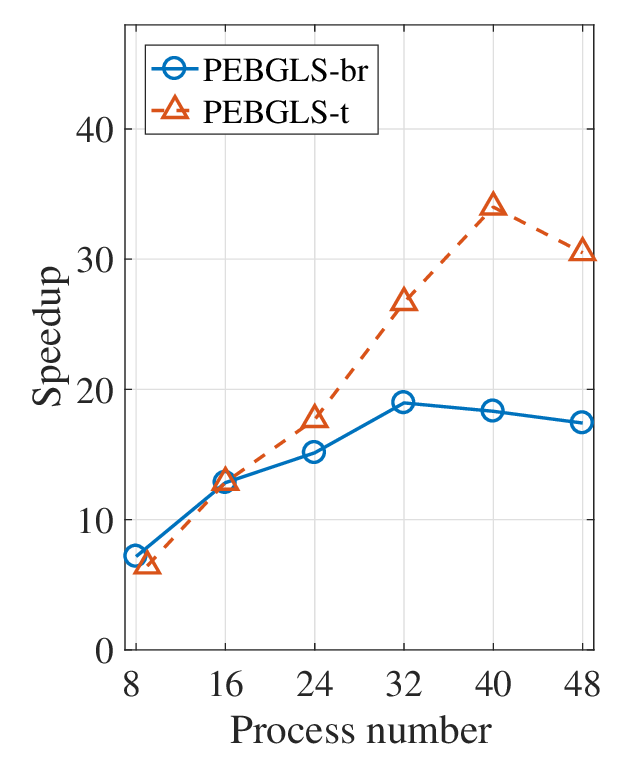}}
  \hspace{0.00\linewidth}
  \subfigure[$e_K$ on att532]{
    %\label{fig:effi1}
    \includegraphics[width=0.32\linewidth]{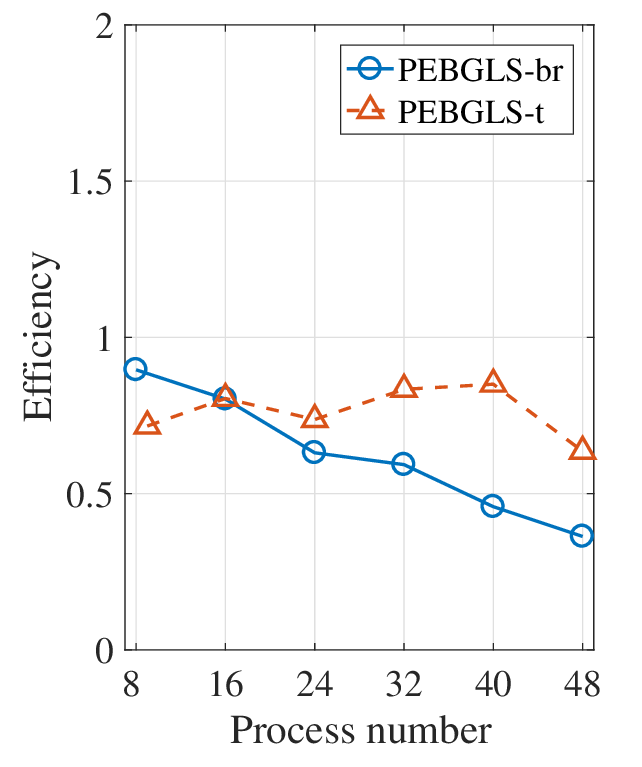}}\\
  \subfigure[$\mathcal{S}_K$ on pr1002]{
    %\label{fig:speedup1}
    \includegraphics[width=0.32\linewidth]{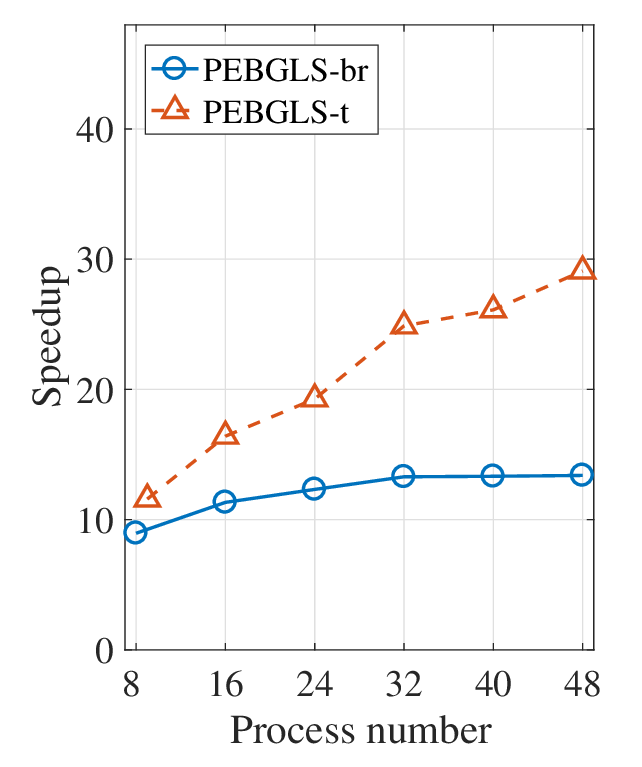}}
  \hspace{0.00\linewidth}
  \subfigure[$e_K$ on pr1002]{
    %\label{fig:effi1}
    \includegraphics[width=0.32\linewidth]{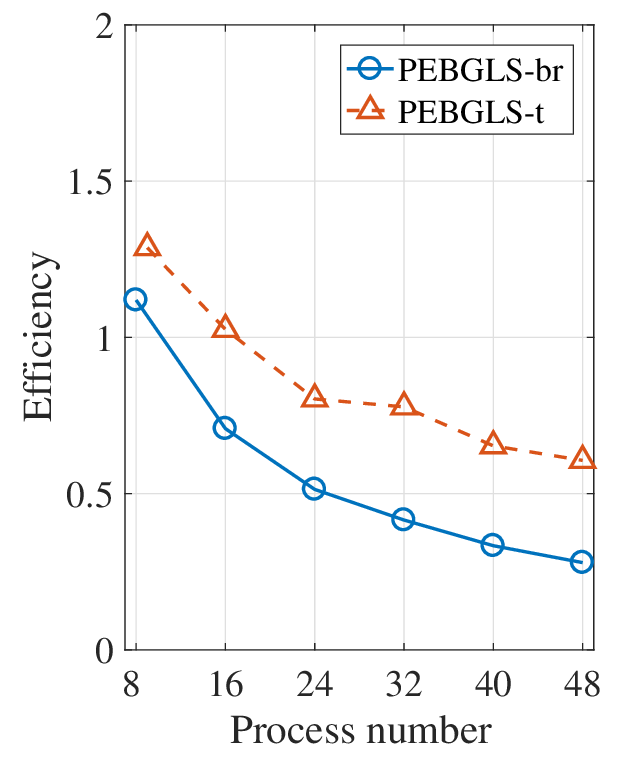}}\\
  \subfigure[$\mathcal{S}_K$ on rl1304]{
    %\label{fig:speedup1}
    \includegraphics[width=0.32\linewidth]{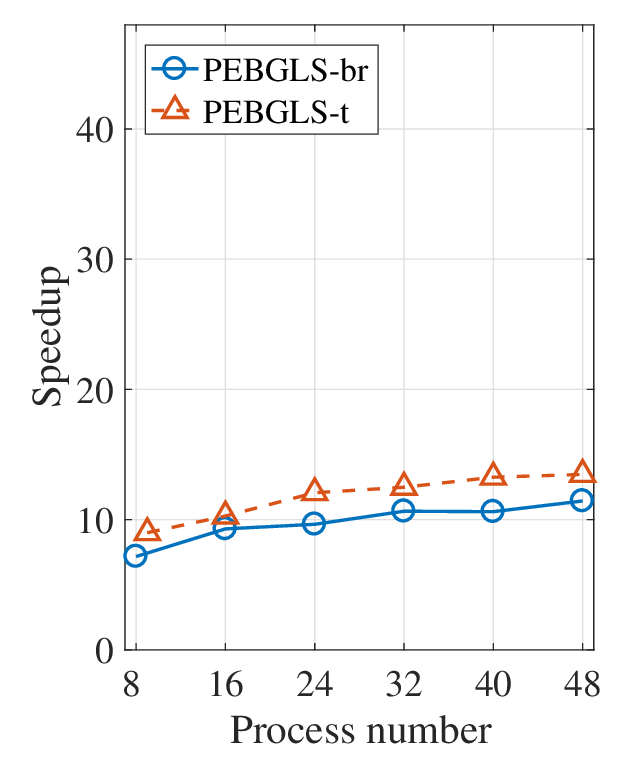}}
  \hspace{0.00\linewidth}
  \subfigure[$e_K$ on rl1304]{
    %\label{fig:effi1}
    \includegraphics[width=0.32\linewidth]{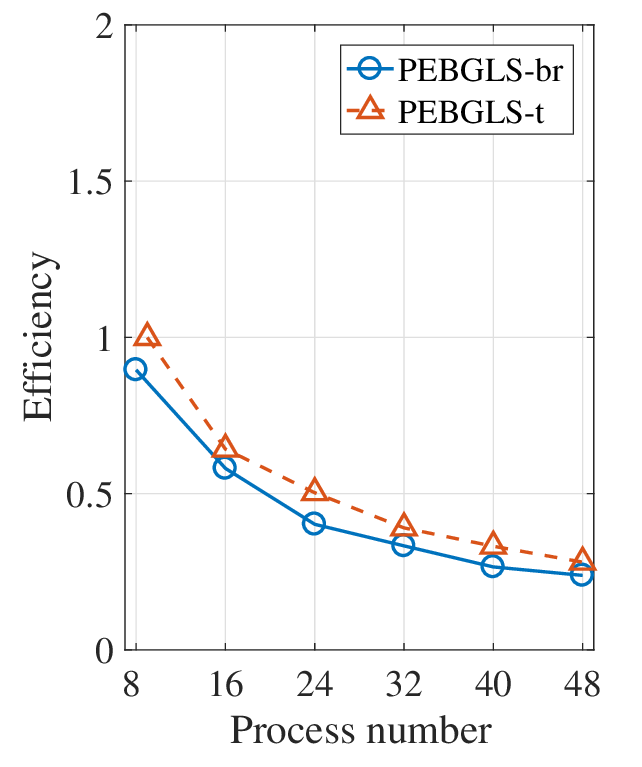}}\\
    %}
  \caption{The speedup $\mathcal{S}_K$ and efficiency $e_K$ of PEBGLS-t and PEBGLS-br}\label{fig:speedup} % epm_201711161653
\end{figure}

The test instances were att532, pr1002 and rl1304 from the TSPLIB~\cite{reinelt1991tsplib}. In TSPLIB, the number in the name of an instance is the node number $n$ of this instance. For PEBGLS, the torus topology (PEBGLS-t) and the bidirectional ring topology (PEBGLS-br) were tested. The experiment was conducted on the Tianhe-2 supercomputer. Tianhe-2 is one of the world's top-ranked supercomputers. It is equipped with 17,920 computer nodes, each comprising two Intel Xeon E5-2692 12C (2.200 GHz) processors. So each node has 24 cores and the system supports elastic parallel computing across nodes. We also used the FLS strategy and Bentley's improvement \cite{bentley1992fast} to enhance the efficiency of the 2-Opt heuristic in PEBGLS. Based on \cite{voudouris1999guided}, the coefficient $\lambda$ is calculated by:
\begin{equation}\label{eq:lambda2}
\lambda = 0.3\cdot \frac{g(\mbox{first local optimum})}{n},
\end{equation}
where $g(\mbox{first local optimum})$ is the function value of the first local optimum visited by GLS and $n$ is the number of cities of the TSP instance. Our pilot experiments showed that PEBGLS is not very sensitive to $w$. So here we set $w=2$. $U$ was set to be 100. Our PEBGLS program was implemented in GNU C++ with O2 optimizing compilation. The OpenMPI library was used as the message passing tool. To calculate $E[T_1]$, \text{100} runs of single-process PEBGLS were executed. The runs started from randomly generated solutions. Because the global optimally costs of the test instances are known, all runs terminated only when the globally optimal cost was achieved. The wall-clock time of each run was recorded. To calculate $E[T_K]$, \text{100} runs of $K$-process PEBGLS were executed on $K$ cores. Each process occupied a core. For PEBGLS-br, $K$ separately took the values $\{8,16,24,32,40,48\}$. For PEBGLS-t, $K$ separately took the values $\{9,16,24,32,40,48\}$ and the shape of the torus topology were \{$(3\times3), (4\times4), (4\times6), (4\times8), (5\times8), (6\times8)$\} respectively. The resulting speedup $\mathcal{S}_K$ values and efficiency $e_K$ values are shown in Table~\ref{tbl:speedup} and Figure~\ref{fig:speedup}. Table~\ref{tbl:speedup} also lists the average runtime $\bar T_K$ for each $K$ value.

\begin{table}%[!t]
\caption{Speedup and efficiency of PEBGLS-t and PEBGLS-br}
\centering
\label{tbl:speedup} % epm_201711161653
\resizebox{\textwidth}{!}{
\begin{tabular}{ >{\raggedright}p{40pt}  >{\centering}p{10pt}  >{\centering}p{40pt} >{\centering}p{40pt} >{\centering}p{40pt} >{\centering}p{40pt} >{\centering}p{40pt} >{\centering}p{40pt} >{\centering\arraybackslash}p{40pt} }
\hline
\multicolumn{9}{c}{PEBGLS-br}\\
\hline
 & & $K=1$ & $K=8$ & $K=16$ & $K=24$ & $K=32$ & $K=40$ & $K=48$ \\
 \hline
\multirow{3}{*}{att532} & $\bar T_K$ & 4.9130s & 0.6854s & 0.3827s & 0.3247s & 0.2591s & 0.2682s & 0.2821s\\
\cline{2-9}
 & $\mathcal{S}_K$ & - & 7.1681 & 12.8377 & 15.1309 & 18.9618 & 18.3184 & 17.4158\\
\cline{2-9}
 & $e_K$ & - & 0.8960 & 0.8024 & 0.6305 & 0.5926 & 0.4580 & 0.3628\\
\hline
\multirow{3}{*}{pr1002} & $\bar T_K$ & 22.5781s & 2.5219s & 1.9939s & 1.8329s & 1.6991s & 1.6943s & 1.6857s\\
\cline{2-9}
 & $\mathcal{S}_K$ & - & 8.9528 & 11.3236 & 12.3182 & 13.2883 & 13.3259 & 13.3939\\
\cline{2-9}
 & $e_K$ & - & 1.1191 & 0.7077 & 0.5133 & 0.4153 & 0.3331 & 0.2790\\
\hline
\multirow{3}{*}{rl1304} & $\bar T_K$ & 19.7827s & 2.7610s & 2.1300s & 2.0523s & 1.8586s & 1.8647s & 1.7307s\\
\cline{2-9}
 & $\mathcal{S}_K$ & - & 7.1650 & 9.2877 & 9.6393 & 10.6439 & 10.6091 & 11.4305\\
\cline{2-9}
 & $e_K$ & - & 0.8956 & 0.5805 & 0.4016 & 0.3326 & 0.2652 & 0.2381\\
\hline
\hline
\multicolumn{9}{c}{PEBGLS-t}\\
\hline
 & & $K=1$ & $K=8$ & $K=16$ & $K=24$ & $K=32$ & $K=40$ & $K=48$ \\
 \hline
\multirow{3}{*}{att532} & $\bar T_K$ & 4.9130s & 0.7623s & 0.3818s & 0.2778s & 0.1842s & 0.1445s & 0.1611s\\
\cline{2-9}
 & $\mathcal{S}_K$ & - & 6.4450 & 12.8680 & 17.6854 & 26.6721 & 34.0000 & 30.4966\\
\cline{2-9}
 & $e_K$ & - & 0.7161 & 0.8042 & 0.7369 & 0.8335 & 0.8500 & 0.6353\\
\hline
\multirow{3}{*}{pr1002} & $\bar T_K$ & 22.5781s & 1.9496s & 1.3762s & 1.1720s & 0.9076s & 0.8648s & 0.7761s\\
\cline{2-9}
 & $\mathcal{S}_K$ & - & 11.5809 & 16.4061 & 19.2646 & 24.8767 & 26.1079 & 29.0917\\
\cline{2-9}
 & $e_K$ & - & 1.2868 & 1.0254 & 0.8027 & 0.7774 & 0.6527 & 0.6061\\
\hline
\multirow{3}{*}{rl1304} & $\bar T_K$ & 19.7827s & 2.1997s & 1.9255s & 1.6398s & 1.5846s & 1.4923s & 1.4689s\\
\cline{2-9}
 & $\mathcal{S}_K$ & - & 8.9934 & 10.2741 & 12.0641 & 12.4843 & 13.2565 & 13.4677\\
\cline{2-9}
 & $e_K$ & - & 0.9993 & 0.6421 & 0.5027 & 0.3901 & 0.3314 & 0.2806\\
\hline
\end{tabular}
}
\end{table}

From Table~\ref{tbl:speedup} and Figure~\ref{fig:speedup} we can see that, overall PEBGLS-t achieved higher speedup values than PEBGLS-br, which means the torus topology is better than the bidirectional ring topology on these instances. However, on the largest instance rl1304, the speedup difference between PEBGLS-br and PEBGLS-t was not significant. As the process number $K$ increased, the efficiency values of PEBGLS-br and PEBGLS-t decreased, except for the PEBGLS-t running on att532. On att532, PEBGLS-t attained the highest efficiency value when $K=40$. When $K$ was relatively large, the efficiency values of PEBGLS-br and PEBGLS-t decreased when the problem size increased. For example, when $K=40$, PEBGLS-br attained the highest efficiency value on att532 and the lowest on rl1304, so did PEBGLS-t. On the other hand, when $K$ was relatively small, the efficiency value seems to be unrelated to the problem size. For example, when $K=9$, on pr1002 the efficient value attained by PEBGLS-t was $1.2868$ (superlinear), meanwhile on att532 the value was $0.7161$ (sublinear).

\subsection{Influence of Communication Frequency}\label{sec:epm_influence_U}
The performance of a parallel metaheuristic is influenced by the communication load among processes. In PEBGLS, every $U$ iterations, each process sends $s_{hb}$ to all neighbors if $s_{hb}$ has changed since the previous sending. Hence the communication frequency of PEBGLS has a negative relation to the parameter $U$. In this section, we conducted an experiment to investigate the influence of communication frequency in PEBGLS by setting different $U$ values.

The platform of this experiment was the Tianhe-2 supercomputer. We selected pr1002, pr2392, fnl4461, rl5915, pla7397 and rl11849 from the TSPLIB as the test instances. For PEBGLS, the torus topology (PEBGLS-t) and the bidirectional ring topology (PEBGLS-br) were tested. The maximum runtime for different instances were different: \{pr1002:21s, pr2392:48s, fnl4461:90s, rl5915:119s, pla7397:237s, rl11849:371s\}. For each TSP instance and each $U$ value, we executed 20 runs for each algorithm. In each run, $K=48$ processes started from different random solutions and the torus topology shape in PEBGLS-t was $6\times8$. If an algorithm run attains the globally optimal cost before the maximum runtime, it will stop immediately. $U$ separately took the values of $\{1, 500, 1000, 2000\}$. The other experimental settings were the same as the settings in Section~\ref{sec:epm_speedup}. The performance metrics are \emph{excess} and \emph{runtime}, in which the excess is defined by:
\begin{equation}\label{eq:excess2}
    \mbox{excess} = \frac{\mbox{solution cost} - \mbox{globally optimal cost}}{\mbox{globally optimal cost}}\times 100\%.
\end{equation}

\begin{figure}
  %\centerline{
  \centering
  \subfigure[Runtime on pr1002]{
    \label{fig:excess_U_pr1002} %% label for first subfigure
    \includegraphics[width=0.47\linewidth]{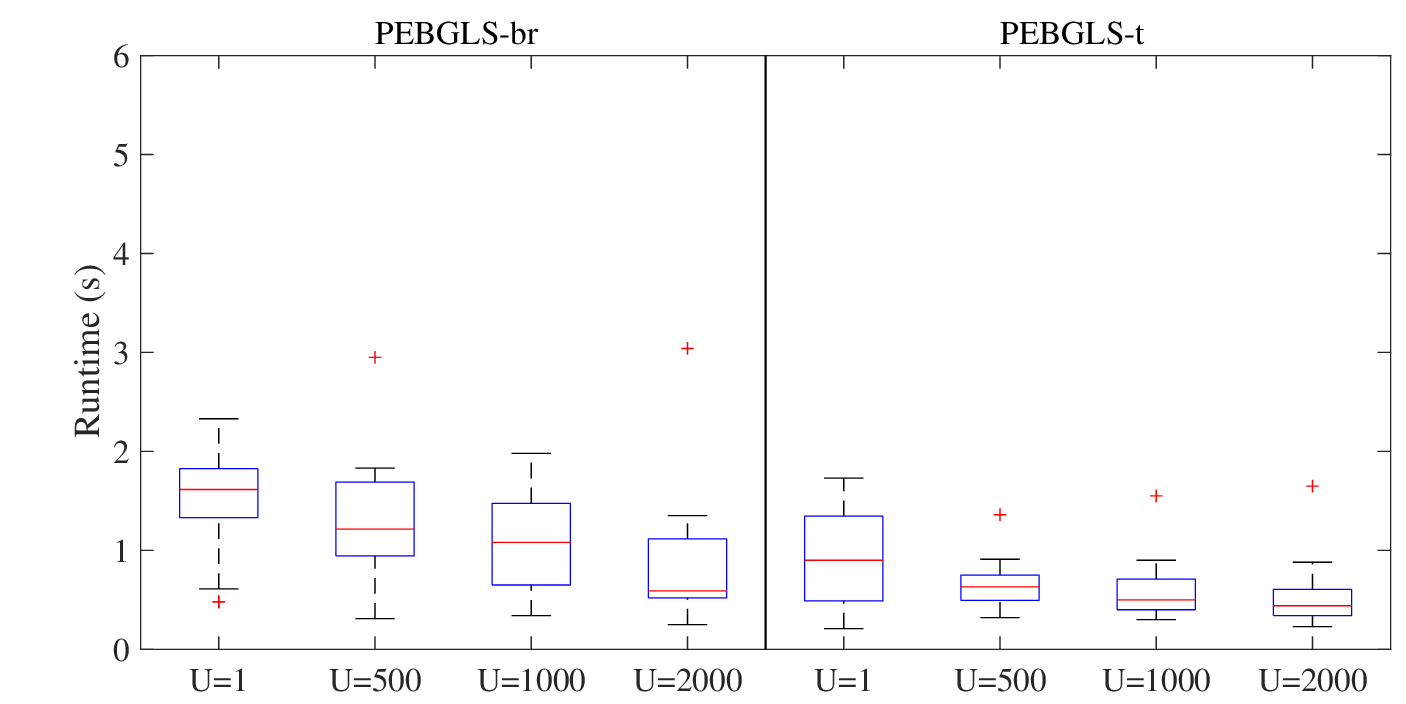}}
    \hspace{0.00\linewidth}
  \subfigure[Runtime on pr2392]{
    \label{fig:excess_U_pr2392} %% label for first subfigure
    \includegraphics[width=0.47\linewidth]{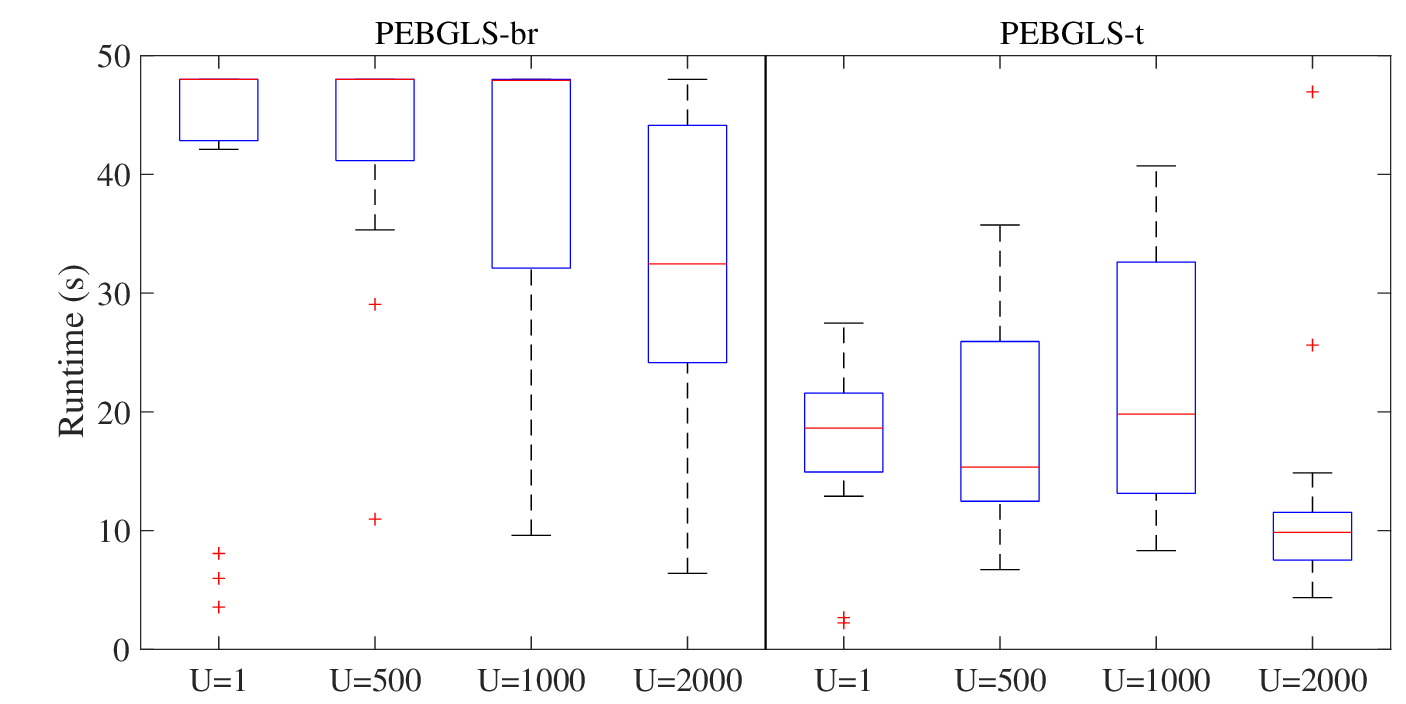}}\\
  \subfigure[Excess on fnl4461]{
    \label{fig:excess_U_fnl4461} %% label for first subfigure
    \includegraphics[width=0.47\linewidth]{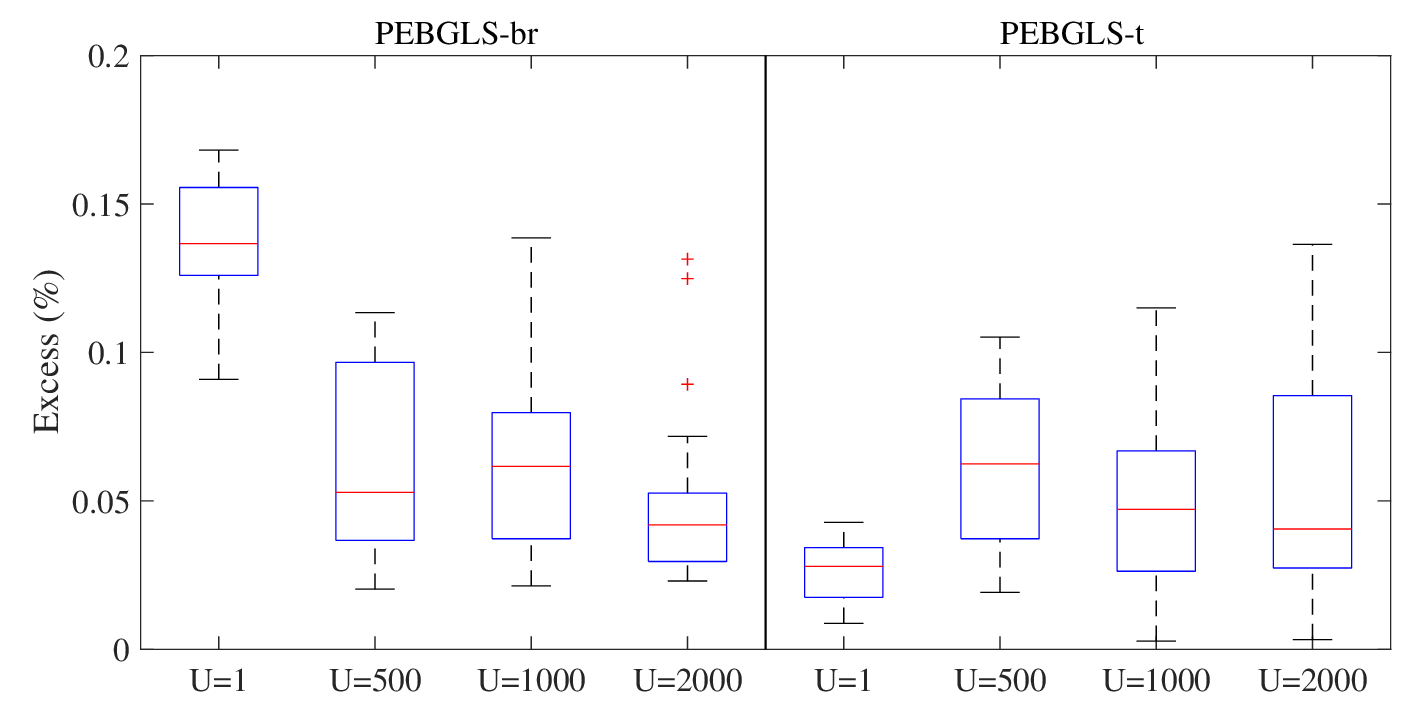}}
    \hspace{0.00\linewidth}
  \subfigure[Excess on rl5915]{
    \label{fig:excess_U_rl5915} %% label for first subfigure
    \includegraphics[width=0.47\linewidth]{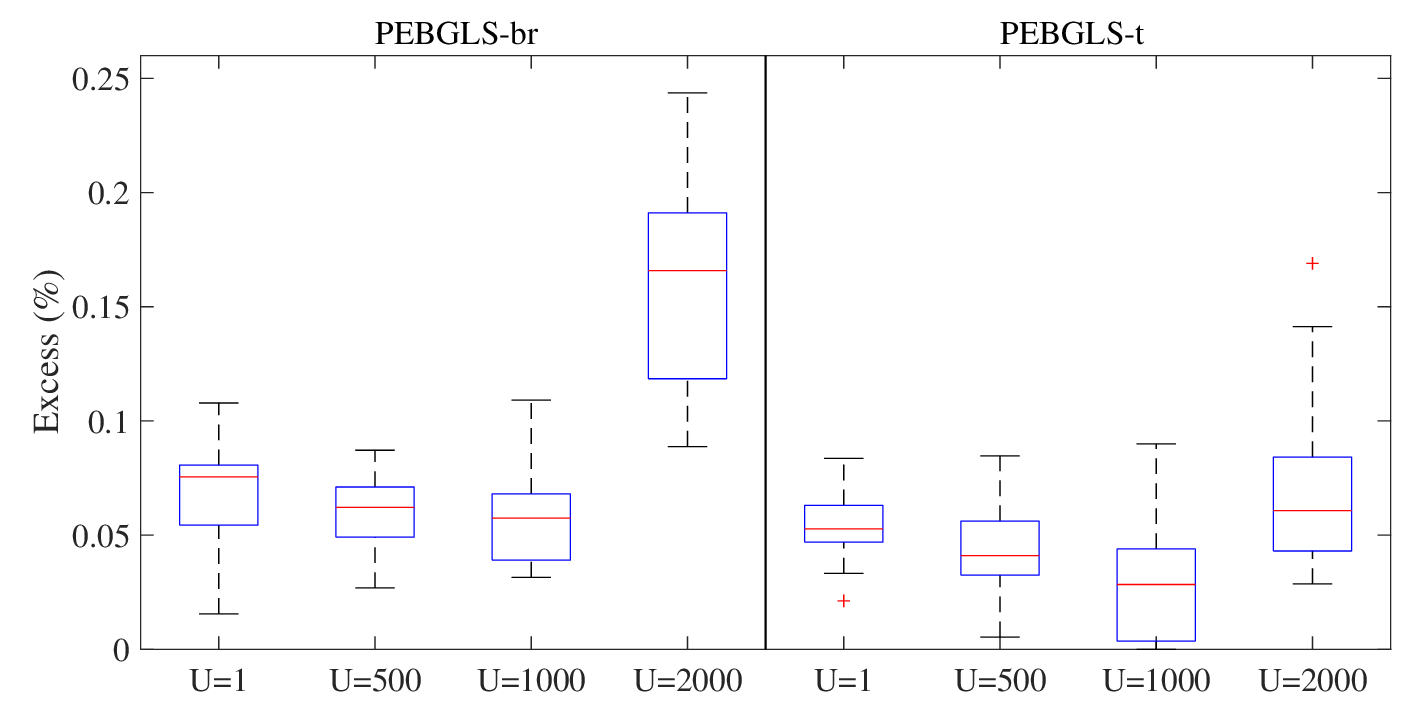}}\\
  \subfigure[Excess on pla7397]{
    \label{fig:excess_U_pla7397} %% label for first subfigure
    \includegraphics[width=0.47\linewidth]{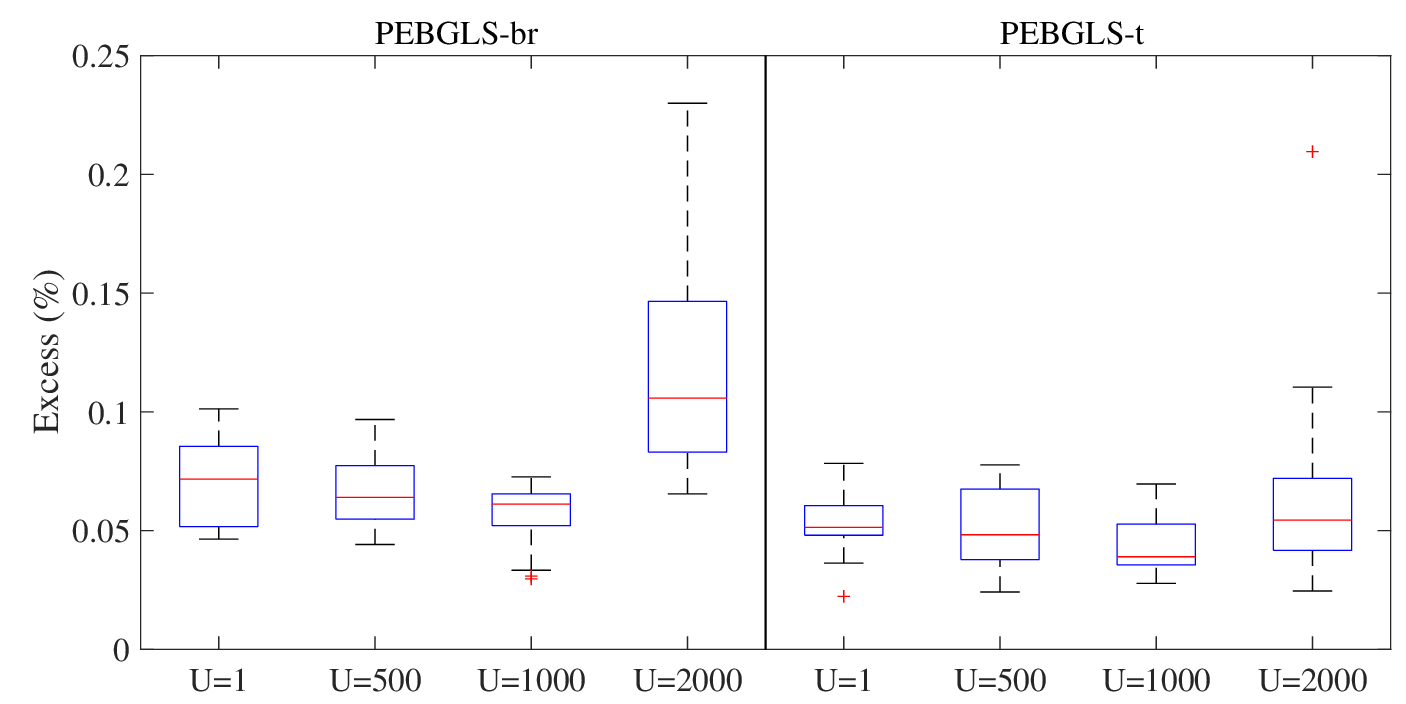}}
    \hspace{0.00\linewidth}
  \subfigure[Excess on rl11849]{
    \label{fig:excess_U_rl11849} %% label for first subfigure
    \includegraphics[width=0.47\linewidth]{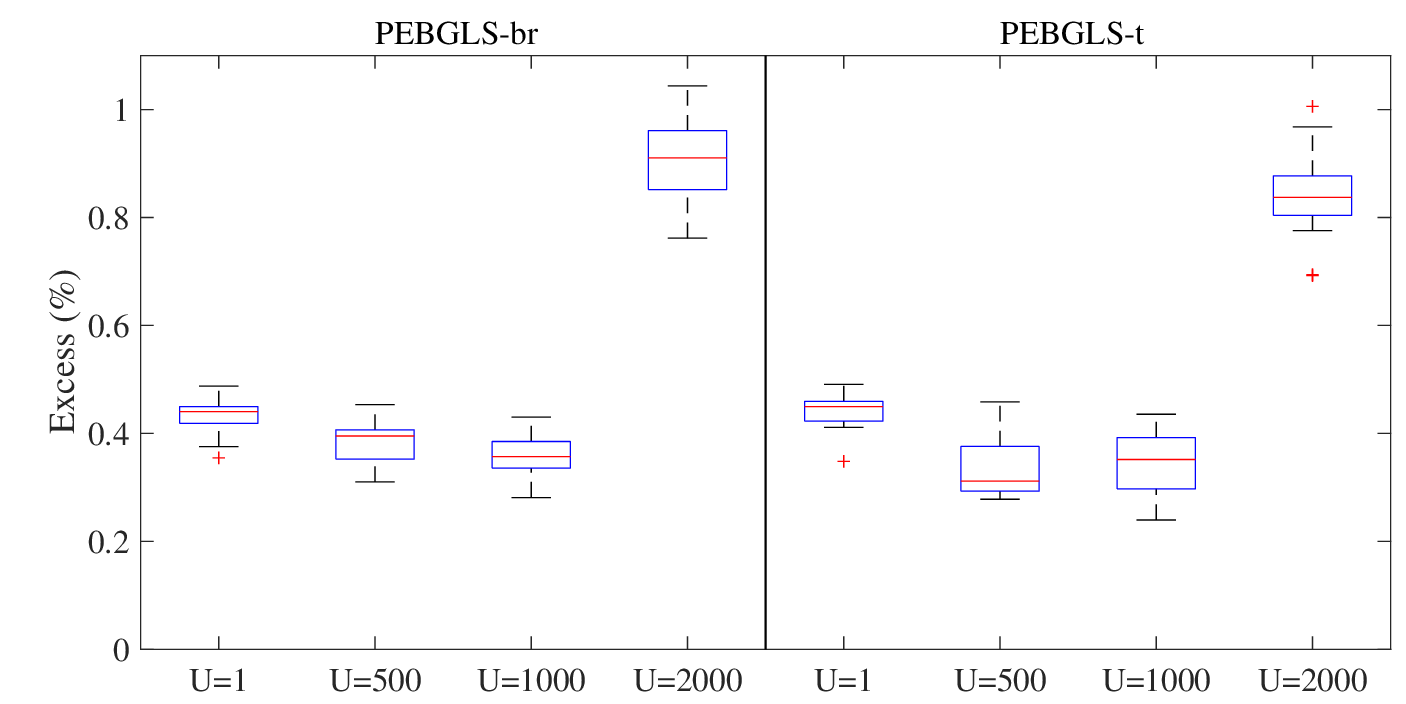}}\\
  %}
  \caption{Excesses and runtime of PEBGLS-br and PEBGLS-t, process number $K=48$}\label{fig:excess_U} % epm_201711151711 epm_201711151448 epm_201607122031 epm_201607042231
\end{figure}

Table~\ref{tbl:excess_U} shows the average excess and average runtime attained by each algorithm, in which the best metric values are in bold. We can see that, on pr1002 and pr2392, most runs of the algorithms attained zero excess, which means that the globally optimal cost was reached before the maximum runtime. Hence in Figure~\ref{fig:excess_U_pr1002} and Figure~\ref{fig:excess_U_pr2392} we present the boxplot of the real runtime of each algorithm on these two instances. On the rest four instances, no algorithm reached the globally optimal cost, hence in Figure~\ref{fig:excess_U_fnl4461}, Figure~\ref{fig:excess_U_rl5915}, Figure~\ref{fig:excess_U_pla7397} and Figure~\ref{fig:excess_U_rl11849} we present the boxplot of the best excess attained by each algorithm.

\begin{table}%[!t]
\caption{Performance of PEBGLS-br and PEBGLS-t with different $U$ values, process number $K=48$}
\centering
\label{tbl:excess_U} % epm_201711151711
\resizebox{1\textwidth}{!}{
\begin{tabular}{c | c c c c | c c c c}%{ p{50pt}  p{67pt}  >{\centering}p{80pt}  >{\centering}p{75pt}  >{\centering}p{75pt}  >{\centering\arraybackslash}p{75pt}}
\hline
& \multicolumn{4}{c}{PEBGLS-br} & \multicolumn{4}{|c}{PEBGLS-t}\\
\cline{2-9}
 & $U$=1 & $U$=500 & $U$=1000 & $U$=2000 & $U$=1 & $U$=500 & $U$=1000 & $U$=2000 \\
 \hline
 Instance & \multicolumn{8}{c}{Average Excess (\%)}\\
 \hline
 pr1002 & 0.0000 & 0.0000 & 0.0000 & 0.0000 & 0.0000 & 0.0000 & 0.0000 & 0.0000 \\
 \hline
pr2392  & 0.0004 & 0.0001 & 0.0001 & 0.0001 & \textbf{0.0000} & \textbf{0.0000} & \textbf{0.0000} & \textbf{0.0000} \\
 \hline
fnl4461  & 0.1368 & 0.0637 & 0.0656 & 0.0506 & \textbf{0.0266} & 0.0613 & 0.0475 & 0.0543 \\
 \hline
rl5915  & 0.0694 & 0.0591 & 0.0578 & 0.1595 & 0.0544 & 0.0432 & \textbf{0.0306} & 0.0693 \\
 \hline
pla7397  & 0.0720 & 0.0658 & 0.0567 & 0.1175 & 0.0534 & 0.0507 & \textbf{0.0445} & 0.0653 \\
 \hline
rl11849  & 0.4311 & 0.3853 & 0.3587 & 0.9066 & 0.4411 & \textbf{0.3366} & 0.3477 & 0.8391 \\
 \hline
 \hline
 Instance & \multicolumn{8}{c}{Average Runtime (s)}\\
 \hline
 pr1002 & 1.51 & 1.33 & 1.07 & 0.85 & 0.90 & 0.65 & 0.59 & \textbf{0.54} \\
 \hline
pr2392 & 40.63 & 42.86 & 39.29 & 31.57 & 17.31 & 18.59 & 22.24 & \textbf{11.87} \\
 \hline
fnl4461 & 90.00 & 90.00 & 90.00 & 90.00 & 90.00 & 90.00 & 90.00 & 90.00 \\
 \hline
rl5915 & 119.00 & 119.00 & 119.00 & 119.00 & 119.00 & 119.00 & 119.00 & 119.00 \\
 \hline
pla7397 & 237.00 & 237.00 & 237.00 & 237.00 & 237.00 & 237.00 & 237.00 & 237.00 \\
 \hline
rl11849 & 371.00 & 371.00 & 371.00 & 371.00 & 371.00 & 371.00 & 371.00 & 371.00 \\
 \hline
\end{tabular}
}
\end{table}

From Table~\ref{tbl:excess_U} and Figure~\ref{fig:excess_U} we can see that, in most cases, PEBGLS-t attained lower excess values or lower runtime than PEBGLS-br, which means that the torus topology is better than the bidirectional ring topology. This conclusion is the same to the conclusion in Section~\ref{sec:epm_speedup}. We also can see that, there was a trade-off between the algorithm performance and the communication frequency (parameter $U$). When the communication frequency was very high (i.e. $U$ was very small), PEBGLS did not perform very well. For example, on rl5915 (Figure~\ref{fig:excess_U_rl5915}) the average excess attained by PEBGLS-t when $U=1$ was worse than when $U=500$ and $1000$. This is because when the communication frequency was very high, each PEBGLS process spent a lot of additional time to communicate with other processes. On the other hand, a relatively low communication frequency also reduced the algorithm performance. For example, on rl11849 (Figure~\ref{fig:excess_U_rl11849}) the average excess attained by PEBGLS-t deteriorated when $U=2000$. An interesting phenomenon is that, on pr1002 and pr2392, the best $U$ value for PEBGLS-t was $2000$, which was larger than the best $U$ values on other instances. That is because pr1002 and pr2392 are smaller than other instances, hence in one second, PEBGLS-t executed more iterations on pr1002 and pr2392 than on other instances.

\subsection{Internal Behavior of PEBGLS-t}\label{sec:epm_behavior}
The previous experiments show that for PEBGLS the torus topology is better than the bidirectional ring topology, hence in the following experiments we only apply the torus topology to PEBGLS. The pervious experiments also show that the collaboration among different PEBGLS processes improves the overall solution quality. In this section we investigate how the collaboration benefits each process. To answer that question, we recorded and studied the internal behavior of PEBGLS-t during the search. For comparison, we also recorded the behavior of Independent PEBGLS processes (P-I-EBGLS). The only different between PEBGLS-t and P-I-EBGLS is that in P-I-EBGLS the processes do not communicate with each other and $s_e$ is only updated by $s_{hb}$ every $U$ iterations.

The experimental platform was Tianhe2 supercomputer. The test instances were gr431, att532 and rat575 from the TSPLIB. On each instance, we first randomly generated $16\times\text{1,000}$ different initial solutions. Then \text{1,000} runs of PEBGLS-t with 16 processes and \text{1,000} runs of P-I-EBGLS with 16 processes were started from the generated solution set. The torus topology shape in PEBGLS-t was $4\times4$. Hence for each PEBGLS-t run, there was a P-I-EBGLS run starting from the same initial solutions. All the runs ended when the globally optimal cost was reached. In the first \text{1,000} iterations, the PEBGLS-t processes did not communicate with each other, i.e. in the first \text{1,000} iterations the PEBGLS-t processes did not cooperate with each other. The other experimental settings were the same as the settings in Section~\ref{sec:epm_speedup}. In this experiment, the entire history of each process was recorded.

\begin{figure}
  \centering
  \subfigure[gr431]{
  \label{fig:excess_vs_iter_gr431}
  \includegraphics[width=0.7\linewidth]{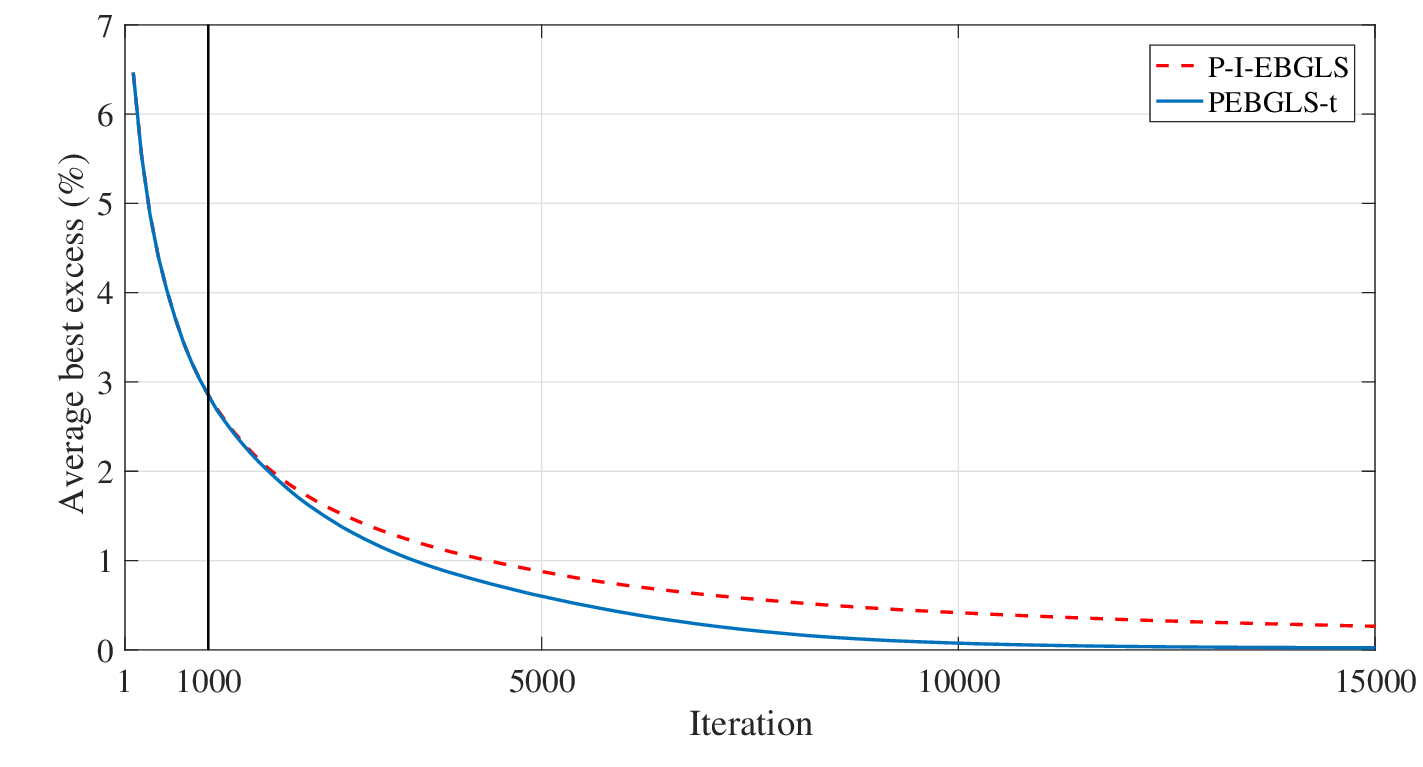}}\\
  \subfigure[att532]{
  \label{fig:excess_vs_iter_att532}
  \includegraphics[width=0.7\linewidth]{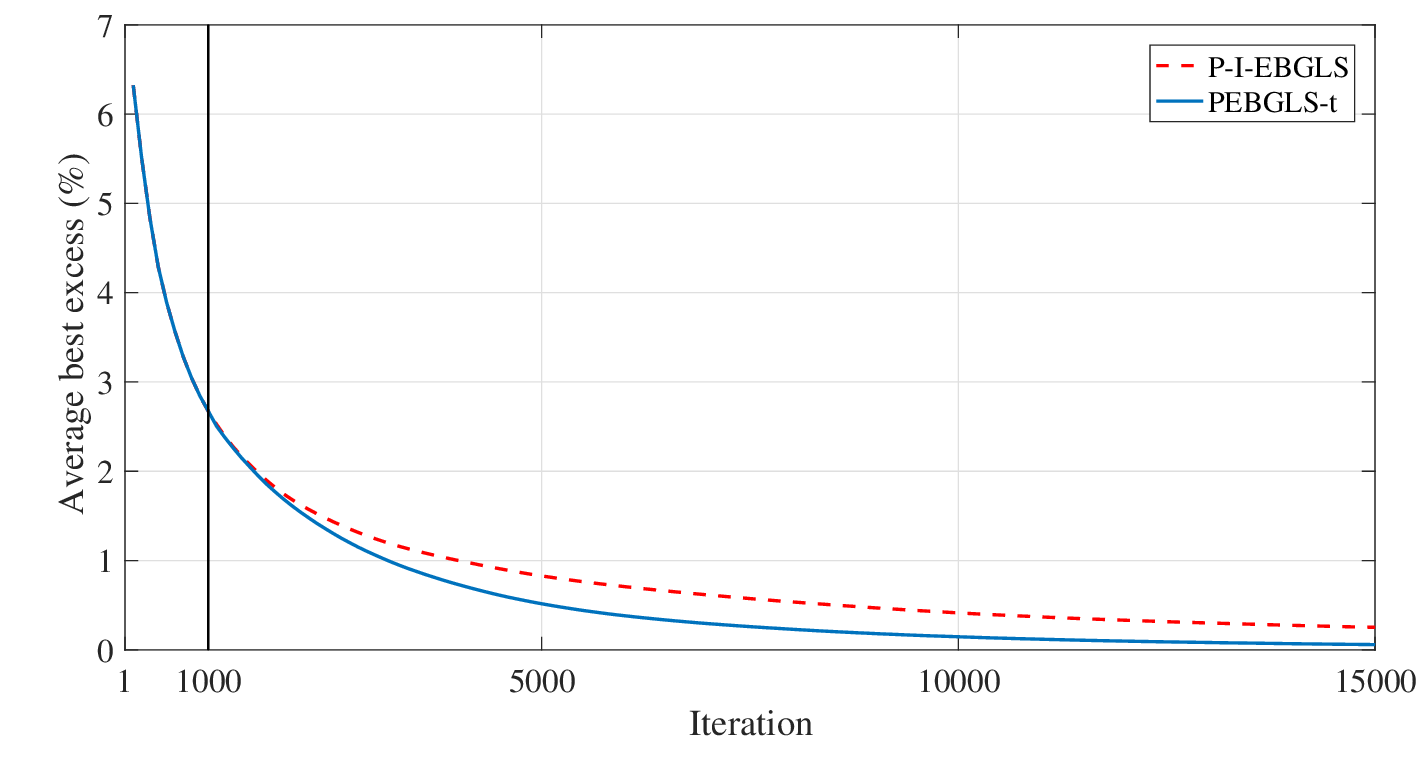}}\\
  \subfigure[rat575]{
  \label{fig:excess_vs_iter_rat575}
  \includegraphics[width=0.7\linewidth]{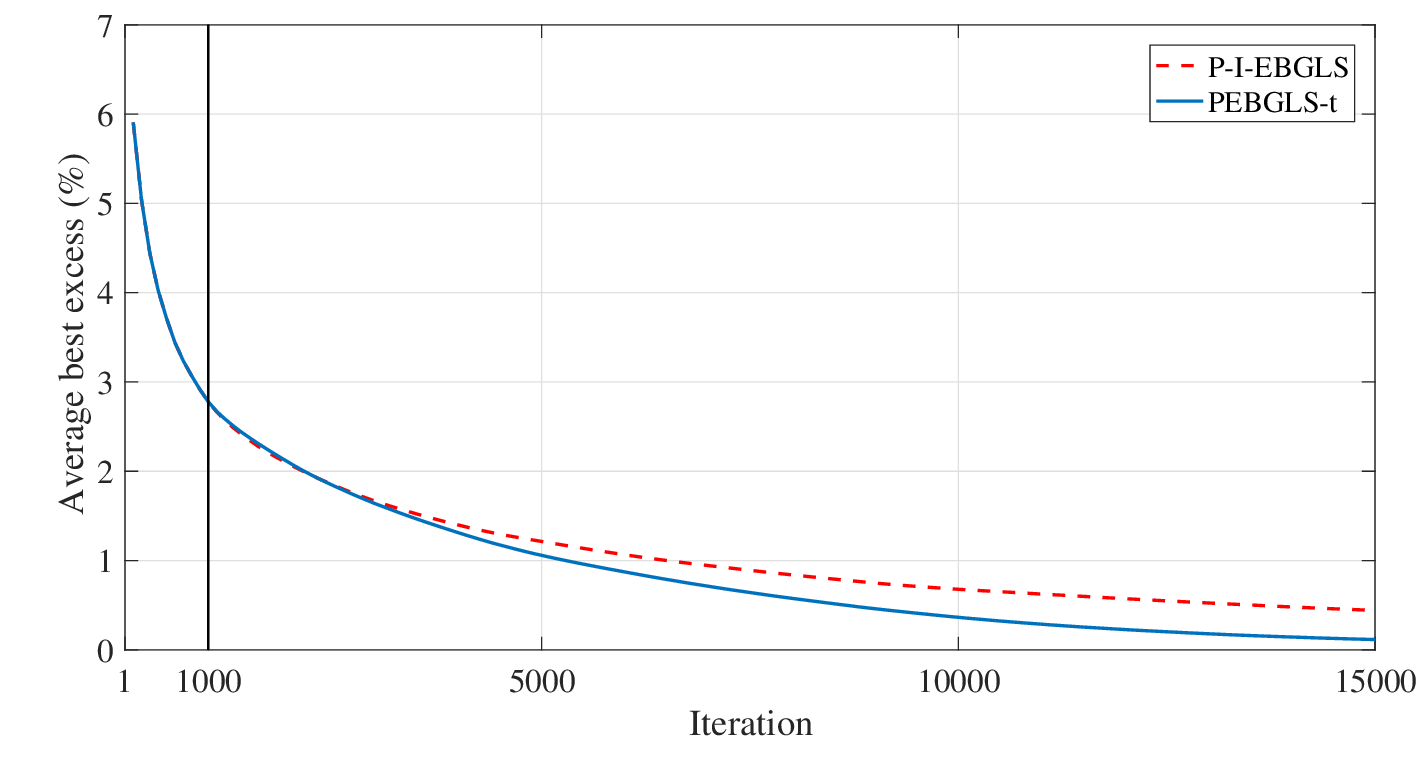}}\\
  \caption{The average best excess versus the time}\label{fig:excess_vs_iter} % epm_201608171519 epm_201711132004 epm_201711141632
\end{figure}

Figure~\ref{fig:excess_vs_iter} shows how the average excess changed over time on the three instances. We can see that in the first \text{1,000} iterations, the average excess attained by PEBGLS-t was the same as that attained by P-I-EBGLS. This is because in the first \text{1,000} iterations the PEBGLS-t processes did not communicate with each other. After the \text{1,000}th iteration, the PEBGLS-t processes started to communicate with each other. Then the average excess attained by PEBGLS-t became lower than that attained by P-I-EBGLS. This means that the cooperation approach in the PEB framework can truly improve the overall solution quality.

After the experiment, we analyzed all the returned solutions of the runs of PEBGLS-t and P-I-EBGLS. After eliminating the duplicated solutions, there were only two unique solutions left. This means that all the runs of PEBGLS-t and P-I-EBGLS ended in two different globally optimal solutions. By comparing the edges in these two globally optimal solutions, we found that the first one only had two different edges to the second one. In other words, there were totally 534 unique edges in the two globally optimal solutions of att532. We denotes these 534 edges as the \emph{optimal edges} of att532. Using the same method, we found the numbers of optimal edges in gr431 and rat575 were 433 and 577 respectively. Since the optimal edges are the edges belongs to the globally optimal solutions, it is undesirable for PEBGLS-t to penalize the optimal edges.

\begin{figure}
  \centering
  \subfigure[gr431]{
  \label{fig:pen_ratio_vs_iter_gr431}
  \includegraphics[width=0.7\linewidth]{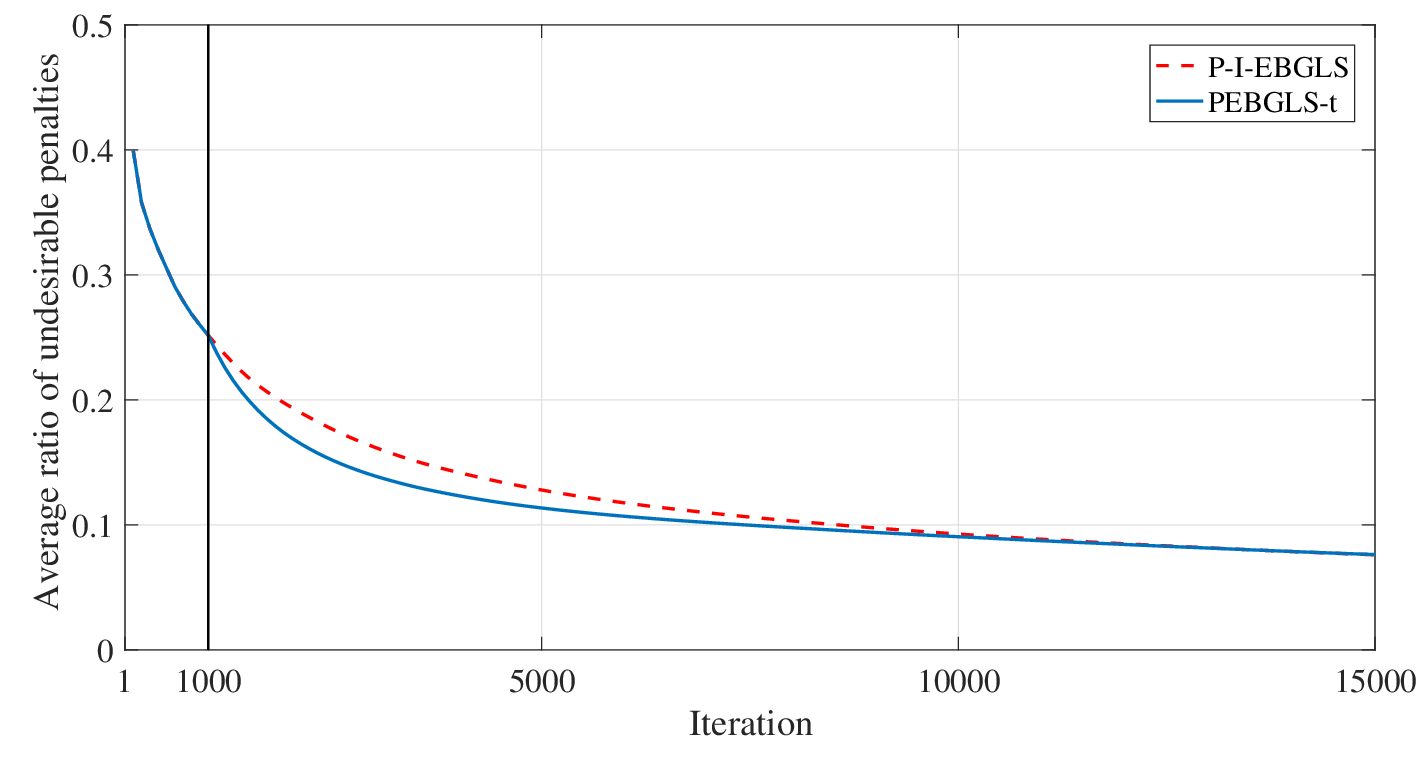}}\\
  \subfigure[att532]{
  \label{fig:pen_ratio_vs_iter_att532}
  \includegraphics[width=0.7\linewidth]{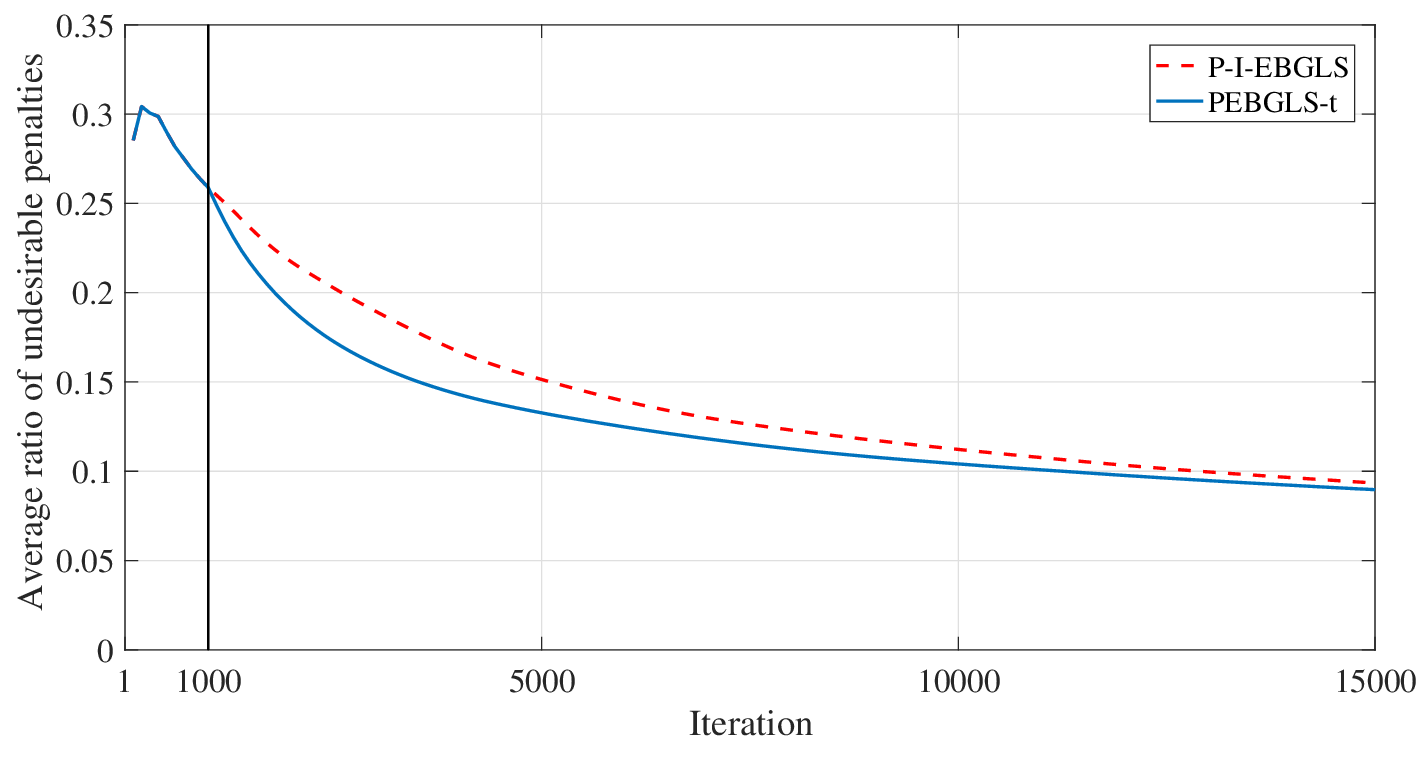}}\\
  \subfigure[rat575]{
  \label{fig:pen_ratio_vs_iter_rat575}
  \includegraphics[width=0.7\linewidth]{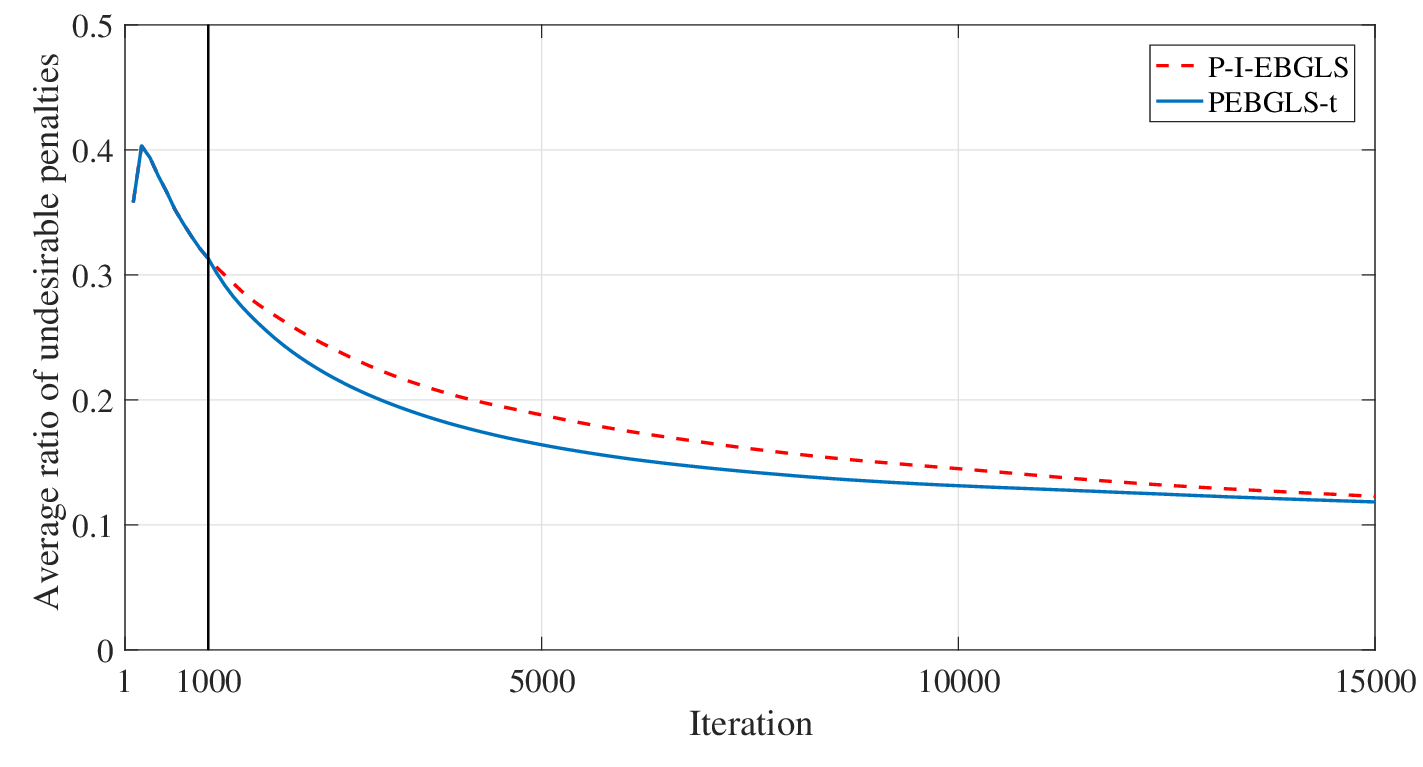}}\\
  \caption{The average ratio of undesirable penalties versus the time}\label{fig:pen_ratio_vs_iter}% epm_201608171519 epm_201711132004 epm_201711141632
\end{figure}

For each search process, we define a metric called \emph{ratio of undesirable penalties}, denoted by $r$, which is the ratio between the total penalty imposed on the optimal edges over the total penalty imposed on all the edges in a TSP instance. We use $E'$ to denote the set of optimal edges. Then $r$ is defined as:
\begin{equation}\label{eq:pen_ratio}
r =\frac{\sum\limits_{i\in E'}p_{i}}{\sum\limits_{i\in E}p_{i}}.
\end{equation}
where $p_{i}$ is the penalty on edge $i$, $E$ is the set of all edges of the TSP instance. Then we calculated the average ratio of undesirable penalties $\bar{r}$ among all processes in all runs. Obviously, everything being equal, a lower ratio value means a more effective penalizing mechanism of PEBGLS-t. Figure \ref{fig:pen_ratio_vs_iter} shows how the average ratio of undesirable penalties changed with the time on these three instances.

In Figure \ref{fig:pen_ratio_vs_iter}, after the PEBGLS-t processes started to cooperate with each other, their average ratio value became smaller than that of the P-I-EBGLS processes. This means that, sharing elite solutions among processes reduced the probability of penalizing the edges in the globally optimal solutions. Hence the search processes in PEBGLS-t became more targeted and the probability of finding the global optima was increased. According to the ``big valley'' structure~\cite{boese1995cost} of the symmetric TSP, high-quality solutions are more likely to have more common edges with the global optima. Hence by reducing the penalties imposed on the edges of the global optima, the PEBGLS-t processes had more chance to find high-quality solutions. However, in Figure \ref{fig:pen_ratio_vs_iter} we can also see that the difference between the two curves became smaller at the final stage of the search. This is because GLS only increases the penalties imposed on the local optima it finds. As illustrated in Figure \ref{fig:excess_vs_iter}, the PEBGLS-t processes found better local optima compared to the P-I-EBGLS processes. According to the big valley structure, the local optima found by PEBGLS-t have more common edges with the global optima than the local optima found by P-I-EBGLS. So the difficulty of PEBGLS-t not penalizing the edges of global optima increased. Hence the difference between the ratio value of PEBGLS-t and the ratio value of P-I-EBGLS became smaller as time went by.

\begin{figure}
  %\centerline{
  \centering
  \subfigure[P-I-EBGLS on gr431]{
    \label{fig:contrib_num_i_gr431} %% label for first subfigure
    \includegraphics[width=0.47\linewidth]{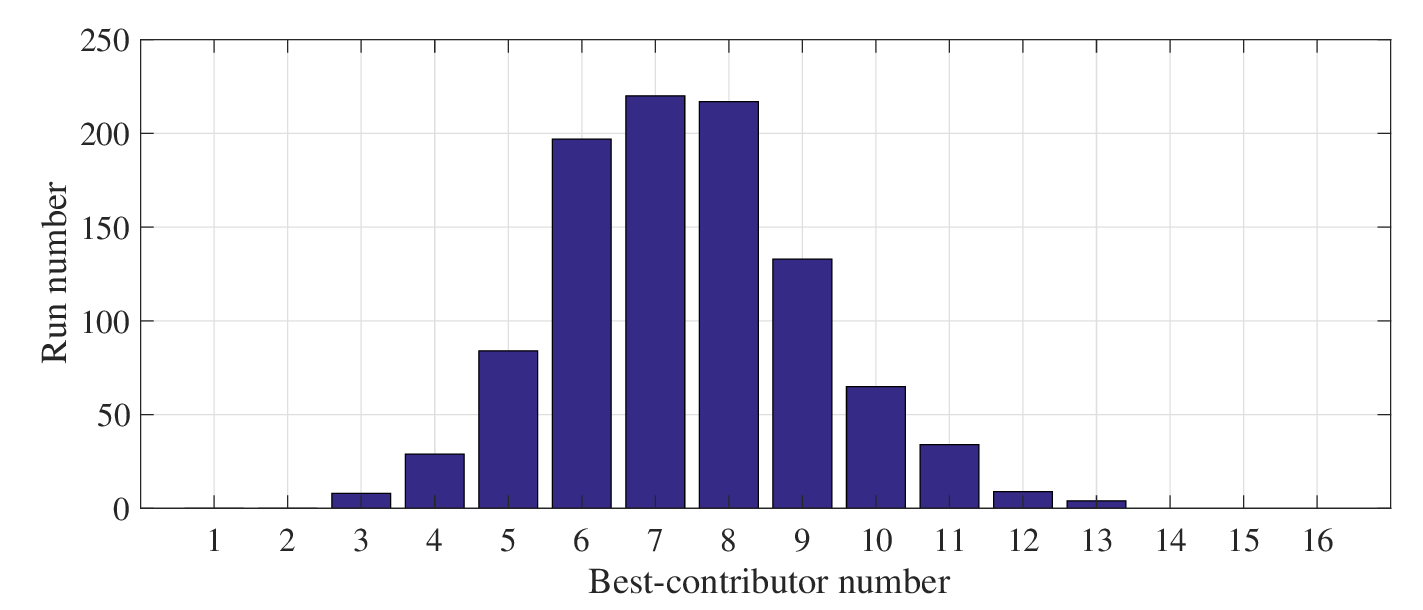}}
    \hspace{0.00\linewidth}
  \subfigure[PEBGLS-t on gr431]{
    \label{fig:contrib_num_t_gr431} %% label for first subfigure
    \includegraphics[width=0.47\linewidth]{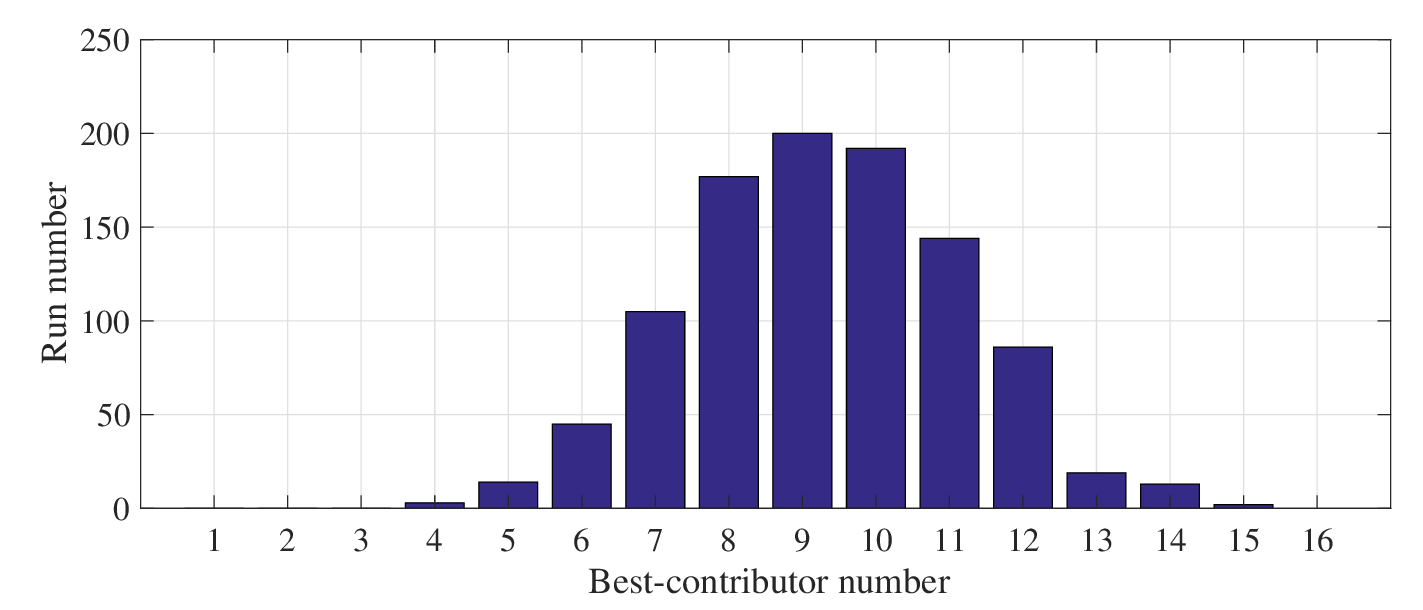}}\\
  \subfigure[P-I-EBGLS on att532]{
    \label{fig:contrib_num_i_att532} %% label for first subfigure
    \includegraphics[width=0.47\linewidth]{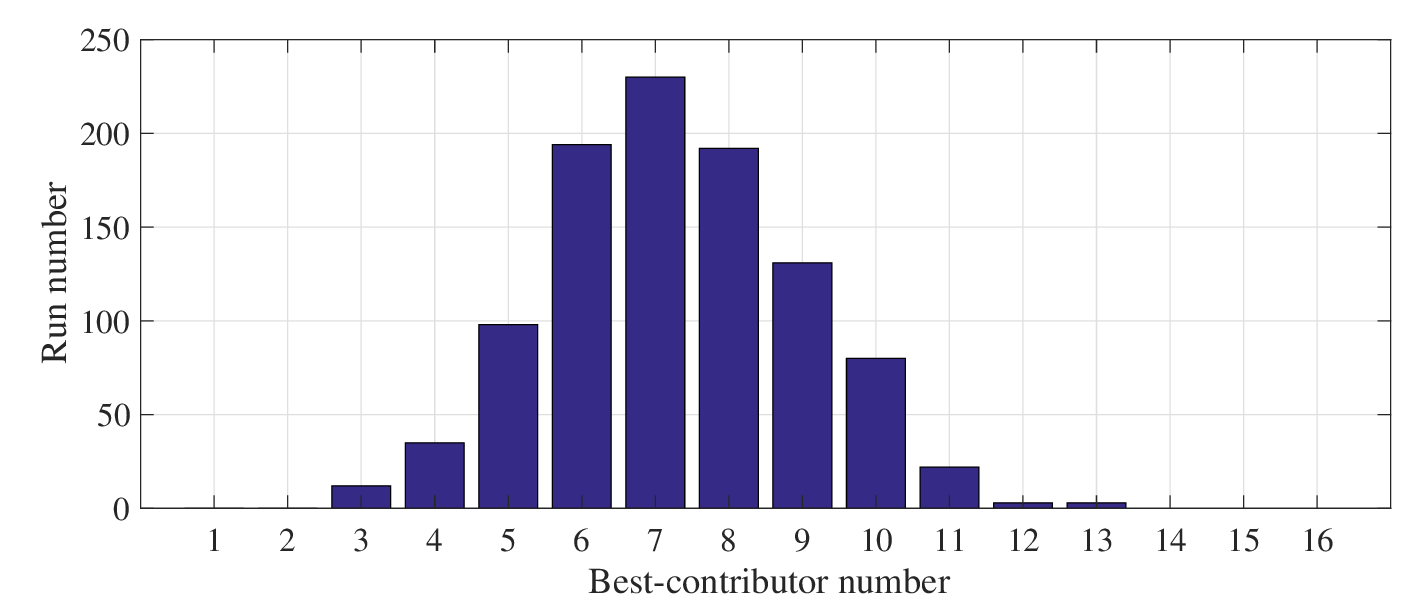}}
    \hspace{0.00\linewidth}
  \subfigure[PEBGLS-t on att532]{
    \label{fig:contrib_num_t_att532} %% label for first subfigure
    \includegraphics[width=0.47\linewidth]{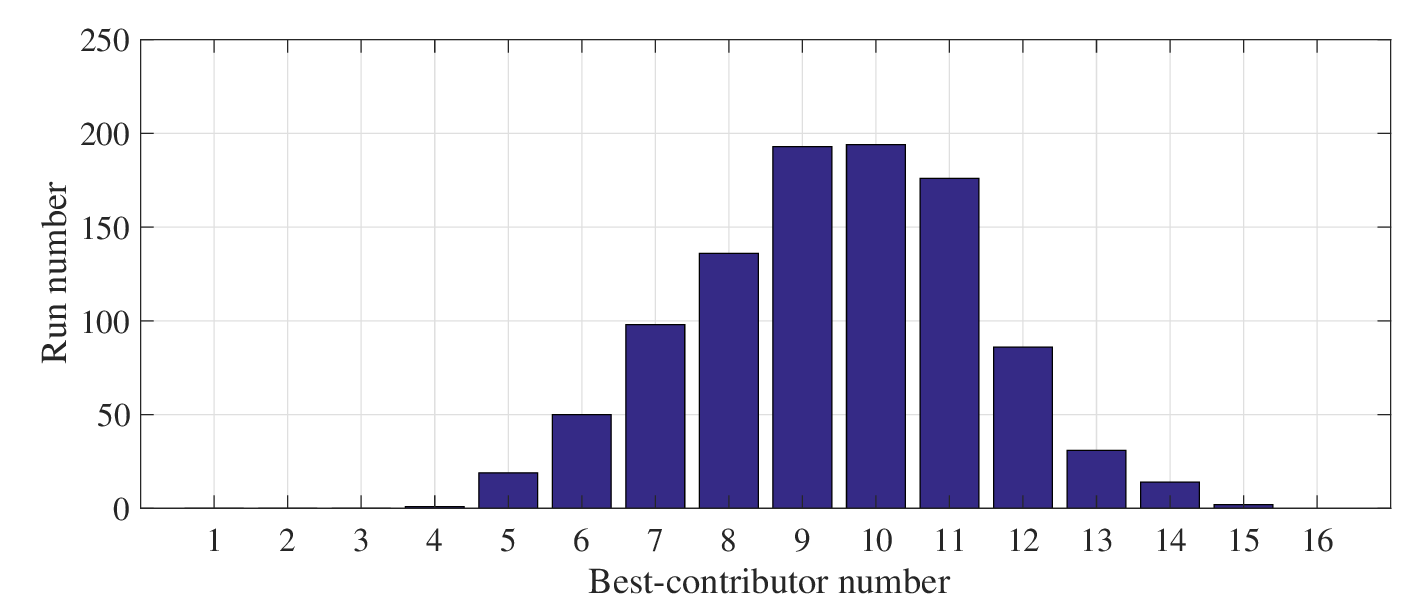}}\\
  \subfigure[P-I-EBGLS on rat575]{
    \label{fig:contrib_num_i_rat575} %% label for first subfigure
    \includegraphics[width=0.47\linewidth]{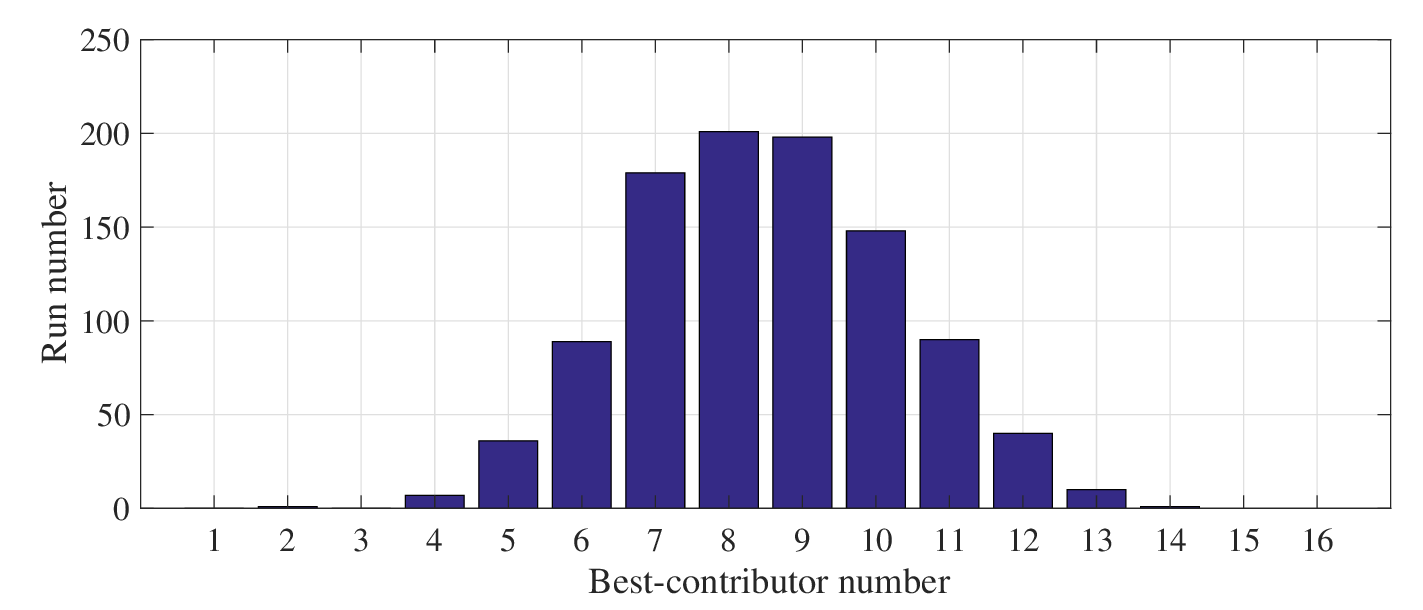}}
    \hspace{0.00\linewidth}
  \subfigure[PEBGLS-t on rat575]{
    \label{fig:contrib_num_t_rat575} %% label for first subfigure
    \includegraphics[width=0.47\linewidth]{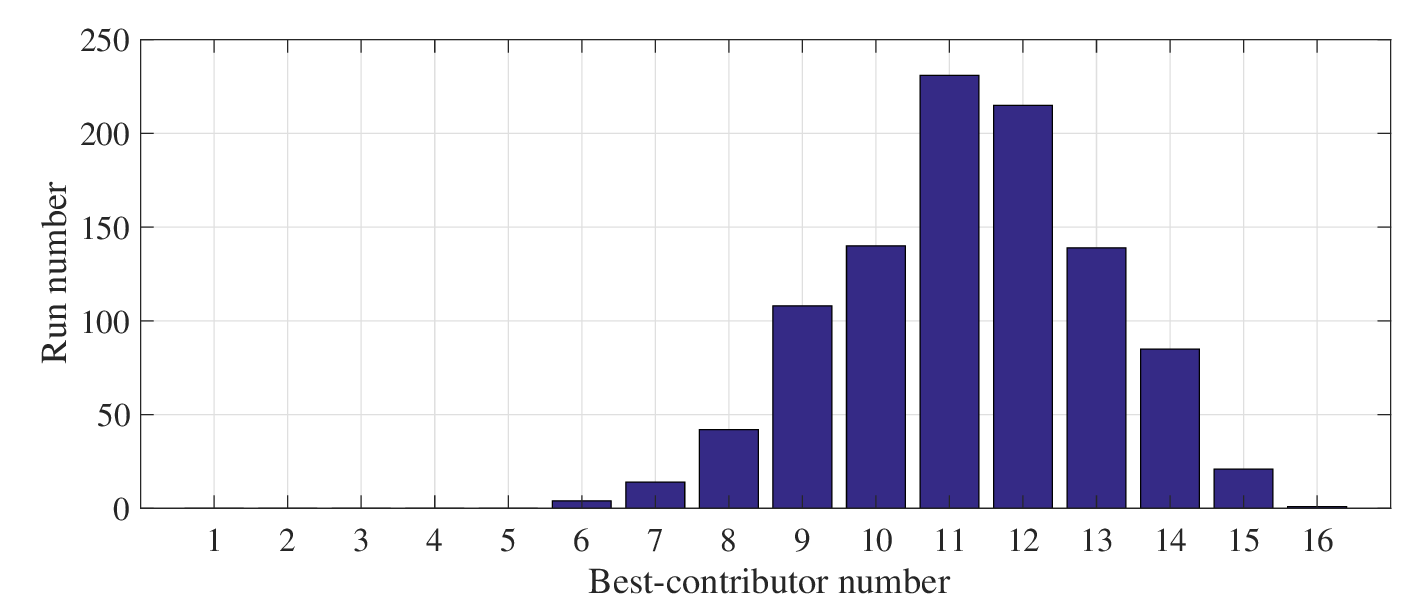}}\\
    %}
  \caption{The distribution of the best-contributor number attained by the \text{1,000} runs of P-I-EBGLS and PEBGLS-t}\label{fig:contrib_num} % epm_201703271505 epm_201711132004 epm_201711141632
\end{figure}

In a parallel trajectory-based metaheuristic, different processes search different regions of the solution space. If a process searches in a less-promising region, this process will contribute little to the global search. We call a process the \emph{best-contributor} if it ever found a solution that was better than the overall best solution among all processes in its search history, i.e., this process once updated the overall best solution. In our experiment, each run of PEBGLS-t/P-I-EBGLS had a certain number of best-contributors. Obviously the number of best-contributors reflects the number of ``useful'' search processes and the overall ``activeness'' of the search processes. Table~\ref{tbl:avg_contrib} shows the average best-contributor number attained by P-I-EBGLS and PEBGLS-t. Figure~\ref{fig:contrib_num} shows the distribution of the best-contributor number on the \text{1,000} runs of P-I-EBGLS and PEBGLS-t. From Table~\ref{tbl:avg_contrib} and Figure~\ref{fig:contrib_num} we can see that, PEBGLS-t had a higher best-contributor number than P-I-EBGLS, which means that the cooperation method in PEBGLS-t increased the overall activity of the processes.

\begin{table}%[!t]
\caption{Average number of best-contributors, for PEBGLS-t and P-I-EBGLS}
\centering
\label{tbl:avg_contrib} % epm_201711141632 epm_201603271505 epm_201711132004
%\resizebox{1\textwidth}{!}{
\begin{tabular}{c c c}%{ p{50pt}  p{67pt}  >{\centering}p{80pt}  >{\centering}p{75pt}  >{\centering}p{75pt}  >{\centering\arraybackslash}p{75pt}}
\hline
& P-I-EBGLS & PEBGLS-t\\
\hline
gr431 & 7.40 & 9.30 \\
att532 & 7.27 & 9.45 \\
rat575 & 8.48 & 11.28 \\
 \hline
\end{tabular}
%}
\end{table}

Based on the above experimental studies we can state that the new cooperation method in PEBGLS-t is effective. This means that the PEB framework can further improve the performance of GLS compared to simply running multiple processes in parallel.

\subsection{Comparison with Other Parallel Metaheuristics}
To test whether the proposed PEBGLS is a competitive TSP metaheuristic, in this section we compared the proposed PEBGLS-t with four parallel metaheuristics for the TSP. The first two are two parallel variants of GLS following the restart based framework. The last two are the parallel variants of two widely-used TSP metaheuristics.

The first comparison algorithm is called Parallel Restart GLS with torus topology (P-R-GLS-t). P-R-GLS-t follows a torus neighborhood topology and executes $K$ GLS processes simultaneously. Every $U$ iterations, each P-R-GLS-t process exchanges its historical best solution $s_{hb}$ with its neighbors. After that, the P-R-GLS-t process abandons its current solution and restarts from the best solution of the set $S_r \cup \{s_{hb}\}$, where $S_r$ is the set of received solutions. The second comparison algorithm combines the PEB framework with the restart based framework, which is called Parallel Restart Elite Biased GLS with torus topology (P-R-EBGLS-t). Every $U$ iterations, each P-R-EBGLS-t process exchanges $s_{hb}$ with its neighbors and restarts from the best solution of the set $S_r \cup \{s_{hb}\}$. Meanwhile, each P-R-EBGLS-t process selects the second best solution from the set $S_r \cup \{s_{hb}\}$ as the elite solution $s_e$. Similar to PEBGLS-t, P-R-EBGLS-t uses the new formula (\ref{eq:modified_util}) to calculate $util$, so that the search direction is attracted by $s_e$.

The third comparison algorithm is a parallel variant of Ant Colony Optimization (P-ACO). We used the ACOTSP software package available at http://www.aco-metaheuristic.org/aco-code/. In P-ACO, $K$ independent ACO processes are executed simultaneously. P-ACO stops when the maximum runtime is achieved or one of the processes finds the globally optimal solution. The forth comparison algorithm is a parallel variant of Iterated Lin-Kernighan algorithm (P-ILK). In our experiment, the ILK implementation came from the Concorde software package available at http://www.math.uwaterloo.ca/tsp/concorde/. The parallelization method of P-ILK was the same to that of P-ACO: $K$ independent ILK processes are executed simultaneously and stop when the maximum runtime is reached or the globally optimal solution is found.

In our experiment, the test TSPLIB instances and the corresponding maximum runtime were \{rd400:8s, att532:11s, gr666:14s, u724:15s, pr1002:21s, d1291:26s, u1432:29s, u1817:37s, pr2392:48s, fnl4461:90s\}. On each instance, the run number of each algorithm was 20 and the process number $K$ was 9. The torus topology shape was $3\times3$. Here we selected a relatively small $K$ value because we intend to show that our algorithm can perform well in multi-core personal computers or small computer clusters. For PEBGLS-t, P-R-GLS-t and P-R-EBGLS-t, we set $U=1000$ and the other settings were the same to the PEBGLS settings in Section~\ref{sec:epm_speedup}. In P-ACO, different ACO processes had different random seeds. For each P-ACO process, the parameter settings were based on \cite{oliveira2011detailed} and shown in Table~\ref{tbl:para_ACO}. In P-ILK, different ILK processes started from different randomly generated solutions. Each ILK process followed the default settings of the Concorde software.

\begin{table}
\caption{Parameter settings of each process in P-ACO}
\centering
\label{tbl:para_ACO}
%\resizebox{\linewidth}{!}{
\begin{tabular}{l|l|l}%{| p{30pt} | >{\raggedleft}p{40pt} | >{\raggedleft}p{30pt} | >{\raggedleft}p{30pt} | >{\raggedleft}p{30pt} | >{\raggedleft}p{30pt} | >{\centering}p{30pt} | >{\centering}p{30pt} | >{\centering}p{30pt} | >{\centering}p{30pt} | >{\centering\arraybackslash}p{90pt} |}
\hline
Parameters & Description & Values\\
\hline
 $m_{\scriptscriptstyle{ACO}}$ & Number of ants & 25\\
 $\alpha$ & Influence of pheromone trails & 1 \\
 $\beta$ & Influence of heuristic information & 2 \\
 $\rho$ & Pheromone trail evaporation & 0.2\\
 LS & Local search & 3-Opt\\
 MMAS & MAX-MIN ant system & Apply\\
\hline
\end{tabular}
%}
\end{table}

Table~\ref{tbl:compare_5_alg} shows the comparison results of the five algorithms, in which the best metric values are in bold. From Table~\ref{tbl:compare_5_alg} we can see that, among all the five parallel metaheuristics, PEBGLS-t performed the best on most instances. For example, on the instances u724, PEBGLS-t achieved a zero average excess value while the other algorithms did not, which means that all the 20 runs of PEBGLS-t found the globally optimal solution. By comparing PEBGLS-t with P-R-GLS-t and P-R-EBGLS-t we can see that, overall the performance of PEBGLS-t was better than that of P-R-EBGLS-t and the performance of P-R-EBGLS-t was better than that of P-R-GLS-t. This means that compared to the restart based framework, the proposed PEB framework can further improve the performance of parallel GLS on these test instances. In summary, the experimental results show the effectiveness of the proposed PEB framework.

\begin{table}%[!t]
\caption{Performance of the five comparison parallel metaheuristics, process number $K=9$}
\centering
\label{tbl:compare_5_alg} % epm_201711231756 epm_201711251700
\resizebox{1\textwidth}{!}{
\begin{tabular}{>{\centering}p{50pt} | >{\centering}p{75pt} | >{\centering}p{75pt} | >{\centering}p{75pt} | >{\centering}p{75pt} | >{\centering\arraybackslash}p{75pt} }
\hline
& PEBGLS-t & P-R-GLS-t & P-R-EBGLS-t & P-ACO & P-ILK\\
 \hline
 Instance & \multicolumn{5}{c}{Average Excess (\%)}\\
 \hline
rd400 & \textbf{0.0000} & 0.0023 & 0.0007 & \textbf{0.0000} & \textbf{0.0000} \\
 \hline
att532 & \textbf{0.0000} & 0.0031 & 0.0031 & 0.0038 & \textbf{0.0000} \\%
\hline
gr666 & \textbf{0.0003} & 0.0173 & 0.0111 & 0.0157 & 0.0022 \\%
\hline
u724 & \textbf{0.0000} & 0.0562 & 0.0452 & 0.0094 & 0.0012 \\
 \hline
pr1002 & \textbf{0.0000} & \textbf{0.0000} & \textbf{0.0000} & 0.0096 & \textbf{0.0000} \\
\hline
d1291 & 0.0073 & 0.0644 & 0.0775 & \textbf{0.0021} & 0.0120 \\%
\hline
u1432 & \textbf{0.0000} & 0.2236 & 0.0918 & 0.2104 & \textbf{0.0000} \\%
\hline
u1817 & \textbf{0.0301} & 0.2677 & 0.2753 & 0.1004 & 0.0877 \\
 \hline
pr2392 & \textbf{0.0011} & 0.2376 & 0.2197 & 0.1881 & 0.0020 \\
 \hline
fnl4461 & 0.0922 & 0.3818 & 0.1873 & 1.5910 & \textbf{0.0401} \\
 \hline
 \hline
 Instance & \multicolumn{5}{c}{Average Runtime (s)}\\
 \hline
rd400 & \textbf{0.10} & 2.89 & 0.95 & 1.27 & 0.13 \\
 \hline
 att532 & \textbf{0.24} & 1.36 & 0.87 & 6.37 & 0.57 \\
 \hline
gr666 & \textbf{5.25} & 9.50 & 8.93 & 12.52 & 5.74 \\%
 \hline
u724 & \textbf{1.29} & 15.00 & 10.87 & 9.44 & 6.76 \\
 \hline
pr1002 & 1.62 & 0.30 & 1.61 & 16.52 & \textbf{0.98} \\
 \hline
 d1291 & \textbf{5.33} & 14.16 & 16.36 & 13.76 & 6.07 \\%
 \hline
u1432 & \textbf{3.81} & 29.00 & 27.61 & 29.00 & 4.81 \\%
 \hline
u1817 & \textbf{23.60} & 37.00 & 37.00 & 37.00 & 35.61 \\
 \hline
pr2392 & 39.22 & 48.00 & 48.00 & 48.00 & \textbf{19.90} \\
 \hline
fnl4461 & 90.00 & 90.00 & 90.00 & 90.00 & 90.00 \\
 \hline
\end{tabular}
}
\end{table}

\section{Conclusion}\label{sec:conclusion}
Parallel metaheuristics can exploit the potential computation power of multi-processor systems. This paper proposes the Parallel Elite Biased framework (PEB framework) to design the parallel variants of trajectory-based metaheuristics. The PEB framework applies a distributed topology and an asynchronous communication strategy. More importantly, the PEB framework employs a new cooperative method, which is different from the widely-used cooperative methods including the restart-based method and the path-relinking method. In the PEB framework, multiple search processes start from different initial solutions. After a predefined period of time, each process communicates with its neighbors to update the set formed by the current historical best solutions found by itself and its neighbors. Then the process selects the best solution in the set as the elite solution $s_e$ and its search direction will be attracted by $s_e$.

The PEB framework has successfully been used to design a parallel Guided Local Search (GLS) metaheuristic called Parallel Elite Biased GLS (PEBGLS) for the Traveling Salesman Problem (TSP). We conducted systematic experiments on the Tianhe-2 supercomputer to test the performance of PEBGLS on the TSP. By analyzing the experimental results, we conclude that PEBGLS is a competitive TSP metaheuristic. Hence the proposed PEB framework is useful in designing efficient parallel trajectory-based metaheuristics. Our work provides a new possible way to design parallel trajectory-based metaheuristics.

\medskip

{\small \noindent
\textbf{Acknowledgments.}
The work described in this paper was supported by the National Science Foundation of China under Grant 61473241, and a grant from ANR/RCC Joint Research Scheme sponsored by the Research Grants Council of the Hong Kong Special Administrative Region, China (Project No. A-CityU101/16).
}



%  \bibliographystyle{elsarticle-num}
%  \bibliography{Ref_Shi}


\end{document}